\documentclass[sigconf]{acmart}
\usepackage{graphicx}

\usepackage{amsmath}
\usepackage{graphicx}
\usepackage{color}
\usepackage{amsfonts}
\usepackage{subfigure}
\usepackage{verbatimbox}

\usepackage{wrapfig} %

\usepackage{algorithm, algpseudocode}  %

\algnewcommand{\LINEIF}[2]{%
    \STATE\algorithmicif\ {#1}\ \algorithmicthen\ {#2} \algorithmicend\ \algorithmicif%
} %
\algnewcommand{\IIf}[1]{\State\algorithmicif\ #1\ \algorithmicthen}
\algnewcommand{\EndIIf}{\unskip\ \algorithmicend\ \algorithmicif}

\DeclareMathOperator*{\argmax}{argmax} %

\newcommand{\region}{R}
\newcommand{\regionset}{\mathcal{\region}}
\newcommand{\sdv}{\sigma}
\newcommand{\expectation}{\mathbb{E}}
\newcommand{\uprofit}{\beta}
\newcommand{\nusers}{n}
\newcommand{\opcost}{c}
\newcommand{\domain}{\mathbb{R}}

\newcommand{\perbset}{\mathcal{Z}}
\newcommand{\dpoint}{\mathbf{x}}
\newcommand{\perbpoint}{z}

\newcommand{\privlvl}{\rho}
\newcommand{\sensitivity}{\nu}
\newcommand{\trustlvl}{\zeta}
\newcommand{\priveval}{\lambda}

\newcommand{\price}{q}
\newcommand{\rate}{k}

\newcommand{\ratesdv}{\rate_{\sdv}}
\newcommand{\user}{u}
\newcommand{\userset}{\mathcal{U}}
\newcommand{\totalnusers}{N}
\newcommand{\addednoise}{\eta}

\newcommand{\actspace}{A}

\newcommand{\rw}{r}

\newcommand{\act}{a}

\newcommand{\terminal}{\tau}
\newcommand{\lasttime}{T}

\newcommand{\actvalueopen}{\textit{open}}
\newcommand{\actvaluecancel}{\textit{cancel}}
\newcommand{\seqact}{\textbf{\act}}
\newcommand{\opening}{\textit{o}}
\newcommand{\buying}{\textit{b}}
\newcommand{\actopen}{\act^{\opening}}
\newcommand{\actbuy}{\act^{\buying}}
\newcommand{\eip}{\text{EIP}}
\newcommand{\ip}{\text{IP}}

\newcommand{\privscale}{d}
\newcommand{\regionsize}{L}
\newcommand{\sigmafactor}{k}
\newcommand{\priceincreasefactor}{h}
\newcommand{\profit}{V}

\newcommand{\bestEIPset}{\mathcal{E}}
\newcommand{\eipvalue}{\mathit{v}}

\newcommand{\poiextravalue}{\eipvalue_e}

\copyrightyear{2020} 
\acmYear{2020} 
\setcopyright{acmlicensed}
\acmConference[SIGSPATIAL '20]{28th International Conference on Advances in Geographic Information Systems}{November 3--6, 2020}{Seattle, WA, USA}
\acmBooktitle{28th International Conference on Advances in Geographic Information Systems (SIGSPATIAL '20), November 3--6, 2020, Seattle, WA, USA}
\acmPrice{15.00}
\acmDOI{10.1145/3397536.3422213}
\acmISBN{978-1-4503-8019-5/20/11}

\begin{document}

\title[Spatial Privacy Pricing]{Spatial Privacy Pricing: The Interplay between Privacy, Utility and Price in Geo-Marketplaces}

\author{Kien Nguyen}
\email{kien.nguyen@usc.edu}
\affiliation{%
  \institution{University of Southern California}
  \city{Los Angeles}
  \state{California}
}

\author{John Krumm}
\email{jckrumm@microsoft.com}
\affiliation{%
  \institution{Microsoft Research AI}
  \city{Redmond}
  \state{Washington}
}

\author{Cyrus Shahabi}
\email{shahabi@usc.edu}
\affiliation{%
  \institution{University of Southern California}
  \city{Los Angeles}
  \state{California}
}

\renewcommand{\shortauthors}{Nguyen, et al.}



\settopmatter{printfolios=true} %

\begin{abstract}
A geo-marketplace allows users to be paid for their location data. Users concerned about privacy may want to charge more for data that pinpoints their location accurately, but may charge less for data that is more vague. A buyer would prefer to minimize data costs, but may have to spend more to get the necessary level of accuracy. We call this interplay between privacy, utility, and price \emph{spatial privacy pricing}. We formalize the issues mathematically with an example problem of a buyer deciding whether or not to open a restaurant by purchasing location data to determine if the potential number of customers is sufficient to open. The problem is expressed as a sequential decision making problem, where the buyer first makes a series of decisions about which data to buy and concludes with a decision about opening the restaurant or not. We present two algorithms to solve this problem, including experiments that show they perform better than baselines.
\end{abstract}

\maketitle

\section{Introduction}
\label{sec:intro}

The current business model of many major technology companies is to provide free services to users in return for their data. Often the services and data are based on location. These freely acquired datasets are then used for various targeted advertising purposes, which in turn pay for the services developed and offered by the company. In addition, sometimes the companies simply sell the users' data to third parties and make a profit. These third parties may be interested in gathering information from certain locations. For example, public health authorities can use the data to identify potential pandemic clusters; city authorities are interested in travel patterns during heavy traffic; or advertisers are interested in the popularity of various locations at different times.  The Location Privacy Protection Act of 2012~\cite{congress2012locprivacyact} tried to address this data collecting and sharing practice. It requires any company that obtains location information from a customer's smartphone to get that customer's express consent before collecting the location data, as well as before sharing the location data with third parties.  However, the current practice is that if a user wants to hide her location data from a service provider, she has to turn off her location-detection device and (temporarily) unsubscribe from the service.  

\begin{figure}[htpb]
    \centering
    \includegraphics[width=0.57\linewidth]{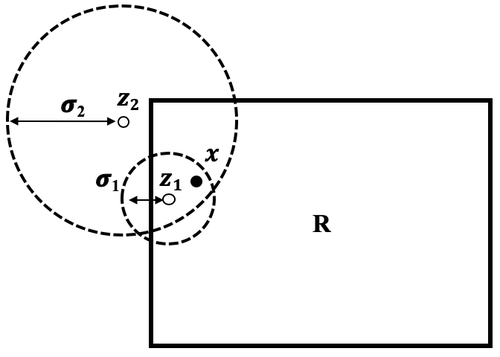}
    \caption{The true location $\dpoint$ (black dot) can be sold as one of the noisy data points $\perbpoint_1, \perbpoint_2$ (white dots). The standard deviation $\sigma_1$ of noise of $\perbpoint_1$ is smaller than that of $\perbpoint_2$. Hence, $\perbpoint_1$ is more expensive than $\perbpoint_2$.}
    \label{fig:example_noisy_points}
\end{figure}

Recently, an alternative framework is emerging to create data marketplaces through which data owners offer their location data to potential buyers~\cite{kanza2015geomarketplace}, dubbed a geo-marketplace~\cite{Kien2019privateGeomarketplace}. Data marketplaces raise a number of interesting issues about data ownership, utility, pricing and privacy. Focusing on geo-marketplaces in this paper, we study the interplay between location data utility, privacy and value (i.e., pricing).  

To illustrate, consider our running example scenario: a buyer is interested in checking if the number of people inside a target region $\region$ is large enough to, say, open a restaurant in $\region$ (utility). 
Determining whether the population is sufficiently large in a given area is also important for a number of decision making applications such as identifying pandemic hotspots, opening COVID-19 test centers, expanding public transportation, opening new community services (e.g. youth centers) and representative government.

Suppose the geo-marketplace has locations of many users, where each user's location can be sold at different levels of accuracy corresponding to her privacy vs. price preferences. 
There are different approaches to capture privacy or inaccuracy. For our purpose, and to simplify the discussion, we assume privacy or inaccuracy is captured as a data point with a Gaussian noise level represented by the standard deviation of the noise distribution.~\footnote{This can be simply extended to use more sophisticated location privacy mechanisms such as Geo-Indistinguishablity~\cite{Andres:2013:geo-indis}.}
In Figure~\ref{fig:example_noisy_points}, for example, a user's true location at $\dpoint$ can be sold as noisy points with mean and standard deviation $(\perbpoint_1, \sigma_1)$ for $\$1$ or $(\perbpoint_2, \sigma_2)$ for $\$2$.

The buyer can purchase data from multiple users and/or multiple times from one user at different levels of accuracy. After making a decision of ``open" the restaurant or ``cancel" (i.e. not open), the buyer would receive a net profit equal to the corresponding revenue of the decision \emph{minus} the cost of purchasing data. The buyer's objective is to maximize this profit.

Maximizing the profit is challenging for the buyer for several reasons. First, the locations of users are uncertain, and the buyer can only reduce this uncertainty by purchasing more data. Second, the purchasing actions are irrevocable (i.e. the buyer cannot ask for reimbursement after already buying the data), so the purchasing action may not be optimal in hindsight. And third, although the problem can be modelled as a partially observable Markov decision process (POMDP), with locations as state and observations and possibly prices as actions, the large number of users and the continuous nature of the domain spaces render the standard POMDP solutions impractical due to the explosion of the number of states.

Some seemingly obvious solutions are not always effective. For example, buying the most accurate data from all users may exceed the payoff for making the correct open/cancel decision, resulting in a negative profit. Alternatively, spending a fixed, prespecified amount on purchasing raises the issue of predetermining that fixed amount. Therefore, there is a need for an adaptive approach to this problem.
Other studies have attempted to find adaptive solutions to similar problems. However, they either considered a simpler version of this problem~\cite{singla2018price} or had different objectives~\cite{gupta2019markovian,ghosh2015selling,nissim2014redrawing,fleischer2012approximatelysellingprivacy}, thus, resulting in inapplicable solutions.
Finally, the approaches for location privacy protection such as spatial cloaking~\cite{vergara2011spatialcloaking}, differential privacy~\cite{dwork2014algorithmic} or Geo-Indistinguishablity~\cite{Andres:2013:geo-indis} are relevant but orthogonal to our work as we discuss in Section~\ref{subsec:privacy_pricing}.
To the best of our knowledge, we are the first to consider the interplay between privacy, utility and price in a data marketplace, particularly in geo-marketplaces, with the focus on the profit of buyers. 

We develop two adaptive algorithms to help buyers optimize the buying actions to obtain necessary data for a decision while striving to reduce the data acquisition cost, called the spatial information probing (SIP) algorithm and the SIP algorithm
with terminals (SIP-T). Our algorithms take into account the uncertainty in the data, the irrevocability of the collection process and the large number of users' possible locations. 
Both algorithms start by buying data at a lower price (dubbed \emph{probing}) in order to gather information, then continue to buy at higher prices the data points that have high potential to give high profit. These algorithms use the \emph{expected incremental profit} ($\eip$), which intuitively is the expected increment of the expected profit when purchasing a data point at a price, to choose the next data point and the next price at which to buy. SIP-T enhances SIP by taking into account the distance of purchased data from the target region and focusing more on distinguishing whether a data point is inside or outside the target region.

Specifically, our contributions are:
\begin{itemize}
    \item Proposing the problem of balancing the benefits for users and buyers in data marketplaces in terms of privacy, utility and price. In the spatial data marketplace, we called this problem \emph{spatial privacy pricing}.
    \item Presenting a specific application in which buyers optimize their purchasing actions to obtain necessary data for a decision while trying to reduce the data acquisition cost.
    \item Developing adaptive algorithms which take into account multiple obstacles such as uncertainty of the data, the irrevocability of the collection process and the large range of possible locations of users. Our algorithms adaptively buy different data points at difference prices based on the $\eip$ and the geometry of the purchased data. 
    \item Extensive experiments comparing our algorithms to baselines over different settings for users' data, buyers' decisions and algorithmic parameters.
\end{itemize}

\section{PROBLEM SETTING}
\label{sec:problem_setting}
In this section, we formalize the problem of the buyer deciding to $\actvalueopen$ or $\actvaluecancel$ a restaurant in a target region $\region$ in order to demonstrate different aspects of the spatial privacy pricing problem: privacy, utility and value (i.e. pricing). We start by defining the notion of privacy valuation of users. We then introduce privacy pricing, which serves as the fundamental mechanism to balance the benefits for users and buyers. Then, we present the decision problem of the buyer, which involves purchasing data from users using the privacy pricing mechanism. Finally, we formalize our problem setting.

\subsection{Users' Privacy Valuation}%
\label{subsec:user_privacy_valuation}
In a data marketplace, the privacy concern or valuation of users can be quantified with three components: their general privacy concern, their concern for a specific data point and their concern about a specific buyer. We present a simple model that captures these three aspects in a straightforward way.

For general privacy valuation, we assume each user $\user_i$ has an overall \emph{privacy level} $\privlvl_i \geq 0$ which reflects the user's own valuation of their privacy and is independent between users. This reflects Westin's series of privacy surveys, where he categorized privacy concerns of people as high, moderate and low~\cite{westin2005survey}.

Each data point $\dpoint_{i, j}$ of user $\user_i$ can have its own \emph{sensitivity} $\sensitivity_{i, j} \geq 0$, which is independent of $\privlvl_i$ and independent between data points. Sensitivity $\sensitivity_{i, j}$ reflects how sensitive $\user_i$ feels about $\dpoint_{i, j}$, e.g. a gas station \emph{vs.} a hospital.

For different buyers, users might also have a different level of \emph{trust} $\trustlvl_b$, e.g., an unknown developer may deserve less trust than the Centers for Disease Control and Prevention.
In this work, since we start with a single buyer with a specific query, we consider $\trustlvl_b = 1$.

Subsequently, the total \emph{privacy valuation} $\priveval_{i, j, b}$ of user $\user_i$ for their data $\dpoint_{i, j}$ for a buyer $b$ would be a function of $\privlvl_i$, $\sensitivity_{i, j}$ and $\trustlvl_b$
\begin{equation}
	\priveval_{i, j, b} = \privlvl_i \sensitivity_{i, j} \trustlvl_b
\end{equation}
When the user has only one data point, we simply use $\priveval_i = \priveval_{i, j, b}$. For simplicity, we use $\priveval_i$ in the remaining discussion.
A user with a higher privacy valuation would expect to receive higher value for their location information.
Next we discuss how $\priveval_i$ affects the price of the users' location data.

\subsection{Privacy Pricing} %
\label{subsec:privacy_pricing}
One popular approach to protect users' privacy when releasing data is by adding noise, either directly to the data or to some components of the data release process. In this work, Gaussian noise is added with the magnitude represented by the standard deviation $\sdv$ dependent on the price paid. The noise magnitude or standard deviation $\sdv$ is considered as the \emph{noise level}.

When a data point $\dpoint_i$ is traded at a price $\price_i$, the noise magnitude $\sdv_i$  is given as a function of $\price_i$ and $\priveval_i$ as follows
\begin{equation}
	\sdv_i = \begin{cases}
	\frac{\priveval_i}{\price_i} \ratesdv &: \quad 0 < \price_i < \priveval_i \\
	0 &: \quad \priveval_i < \price_i
	\end{cases} 
	\label{eq:sdv_from_price}
\end{equation}
where $\ratesdv$ is a scaling factor to make the resulting $\sdv_i$ match with the real-world values of the noise magnitude, discussed in Section~\ref{subsec:param_setup}. This is one possible model that captures the essence of privacy and price.
A higher price $\price_i$ reduces $\sdv_i$, while a higher privacy valuation $\priveval_i$ increases $\sdv_i$.
The privacy valuation $\priveval_i$ can be viewed as the price the user $\user_i$ wants in order to sell their unperturbed data.

Although our users' pricing and privacy models need real-world evaluation, they capture the essence of our model and permit the full development and testing of our proposed algorithms. The functioning of the algorithms is unaffected by the particular pricing and privacy models. For example, other privacy preserving frameworks, such as spatial cloaking~\cite{vergara2011spatialcloaking}, differential privacy~\cite{dwork2014algorithmic} or Geo-Indistinguishability~\cite{Andres:2013:geo-indis}, could also be used instead of the Gaussian noise. In that case, the parameters in those frameworks, e.g. $\epsilon$ in differential privacy, can be derived from a function $\epsilon = f_1(\price_i)$ of the price $\price_i$ with a similar or more complicated pricing model. The noise level $\sdv_i$ then can be derived from a function $\sdv_i = f_2(\epsilon)$ of those parameters. Eventually the noise level $\sdv_i$ can still be considered as a function $\sdv_i = f(\price_i) = f_1\big(f_2(\price_i)\big)$ of the price. The work on specific privacy-preserving techniques or pricing models is orthogonal to our work.
Consequently, in this presentation, we assume the privacy valuation of users and the pricing function are fixed, and we focus on the problem of the buyers maximizing their total reward (i.e. profit) with the strategy to make the final $\actvalueopen$ or $\actvaluecancel$ decision described in the next section.

\subsection{The Buyer's Profit Maximization Problem}
\label{subsec:buyer_decision_making}

In order to finally decide to $\actvalueopen$ or $\actvaluecancel$ (i.e. not open) a restaurant in the target region $\region$, the buyer can take a sequence of actions $\seqact = \{\act_1, \act_2, \dots, \act_{\lasttime}\}$, where taking an action $\act$ would give the buyer a reward or profit $\rw(\act)$.
The goal of the buyer is to maximize the expected total profit of the chosen actions
\begin{equation}
    \max \expectation[\rw(\seqact)] = \max \expectation[\sum_{t = 1}^{\lasttime} \rw(\act_t)]
    \label{eq:max_expected_profit}
\end{equation}
where the expectation is over the (possibly) noisy data from users and the randomness of the process of choosing actions.

There are two types of actions that $\act_t$ can represent: open/cancel $\actopen$ actions and buying $\actbuy$ actions. The buyer can take multiple buying actions to gather information before ultimately deciding to take an open/cancel action. Therefore, the last action $\act_{\lasttime}$ must be an open/cancel action $\actopen$ and other actions $\seqact_{1:{\lasttime}-1} = \{\act_1, \act_2, \dots, \act_{{\lasttime}-1}\}$ must be buying actions $\actbuy$.

An open/cancel action $\actopen$ can take one of the two possible values in the set $\actspace^{\opening} = \{\actvalueopen, \actvaluecancel\}$, which mean open or not open a restaurant, respectively. The profit of an opening action $\actopen$ is
\begin{align}
	\rw(\actopen) = \begin{cases}
	\uprofit \nusers  - \opcost & if \quad \actopen = \actvalueopen \\
	0 & if \quad \actopen = \actvaluecancel
	\end{cases}
	\label{eq:profit_of_opening_action}
\end{align}
where $\uprofit$ is the profit per user (or the gross margin), $\nusers$ is the number of users in the region $\region$ that may visit the restaurant and $\opcost$ is some fixed cost for opening the restaurant, for example, the monthly rent or the cost of operation. Naturally, more users leads to a higher profit. 
This form of the profit function in Equation~\ref{eq:profit_of_opening_action} can capture other variations of the buyer's profit in our problem. For example, the buyer may decide to open only if $\rw(\actvalueopen) > \rw_0$, then $\opcost$ can be set as $\opcost \leftarrow \opcost + \rw_0$. In another example, when the buyer decides to cancel, one may consider the buyer losing a fraction $-\uprofit / k$ of the profit for each user inside $R$ as some opportunity cost. This variation can be captured by setting $\uprofit \leftarrow (1 + \frac{1}{k}) \uprofit$.

Each buying action $\actbuy$ requires the buyer to purchase a data point $\dpoint_i$ at some price $\price > 0$. For clarity, we denote this buying action as $\actbuy(i, \price)$. The set $\actspace^{\buying}$ of all possible buying actions includes all data points at all possible prices. 
The profit for this action $\actbuy(i, \price)$ is the negative of the price the buyer needs to pay, which means
\begin{equation}
    \rw(\actbuy(i, \price)) = -\price
    \label{eq:profit_of_buying_action}
\end{equation}
Thus $\rw(\seqact_{1:\lasttime-1})$ can be considered as the cost of buying data.

In addition to $\rw(\actbuy(i, \price))$, after taking $\actbuy(i, \price)$, the buyer will also receive a noisy data point $\perbpoint_i$ whose coordinates are the true coordinate of $\dpoint_i$ perturbed by independent Gaussian noise with $\sdv_i$ derived from the price $\price$ using Equation~\ref{eq:sdv_from_price}, which means

\begin{equation}
\perbpoint_i = \dpoint_i + \mathbf{\addednoise}, \quad   \mathbf{\addednoise}  \sim \mathcal{N}(\mathbf{0}, \sdv_i^2 I)
\label{eq:noisy_coordinate}
\end{equation}
where $I$ is the $2 \times 2$ identity matrix .

In this work, the number of actions $\lasttime$ the buyer can take is not restricted. In addition, the buyer focuses on their business problem and is honest with the use of data. The buyer is also not restricted by any budget for purchasing data because an excessive cost of purchasing data will eventually decrease the net profit.

\subsection{Formal Definition}
For simplicity, we consider a snapshot in time of users' locations and assume each user has only one data point $\dpoint_i$ at that time with sensitivity $\sensitivity_i$. (Note that this same point could be sold multiple times at different levels of accuracy.) The setting would be similar to the case of multiple data points with minor changes.
With all the quantities defined, we can formalize the problem setting as follows:

Given a snapshot in time of locations of $\totalnusers$ users where each user $u_i$ is at location $\dpoint_i \in \domain^2$, has privacy valuation $\priveval_i$ and is willing to trade $\dpoint_i$ at different prices $\price_i$, the buyer's objective is to maximize the expected total profit of an action sequence $\seqact = \{\act_1, \act_2, \dots, \act_\lasttime\}$ where $\act_\lasttime \in \actspace^{\opening}, \act_i \in \actspace^{\buying} \; \forall i = 1,.., \lasttime-1$:
\begin{equation}
    \max \expectation[\rw(\seqact)] = \max \expectation[\sum_{t = 1}^\lasttime \rw(\act_t)]
    \label{eq:max_expected_profit_repeated}
\end{equation}

\section{Related Work}
\label{sec:background_related_work}

The concept of marketplaces for geosocial data was proposed in~\cite{kanza2015geomarketplace}. In~\cite{aly2019buy}, the authors investigated the value of spatial information to guide the purchases of the buyer. In \cite{Kien2019privateGeomarketplace}, a geo-marketplace was proposed where location data are protected using searchable encryption. However, their setting only consider buying data points at full price, while in our setting one data point can be sold multiple times at different levels of accuracy for different prices. 
Singla~\cite{singla2018price} also studied the problem of maximizing the buyer's profit when information is available at a price, but also can only be purchased at full price, which can be considered as a simpler form of our problem and is used as a baseline in our experiments. In \cite{gupta2019markovian}, Gupta \emph{et. al} extends on the capability of purchasing the same data multiple times. However, they focus on maximizing the total profit, while the objective in our problem is to make a binary decision. In addition, although their method provides some theoretical guarantees of the performance, it relies on the \emph{grades} of states of Markov decision processes that are infeasible to compute in our problem.

A partially observable Markov decision process (POMDP) is a powerful framework for modelling sequential decision making processes under uncertainty with the goal of maximizing total reward. POMDPs have been an active research area and considerable progress has been made on solving POMDPs~\cite{sunberg2018onlinepomdpcontinous,seiler2015online,couetoux2011continuous}. 
While our problem setting can be considered as a POMDP with continuous state, observation and action spaces, our problem also includes a large number of users. Since the number of possible states in this POMDP increases exponentially with the number of users, standard POMDP algorithms are computationally infeasible for our problem.

There are also several studies on selling privacy~\cite{ghosh2015selling,nissim2014redrawing,fleischer2012approximatelysellingprivacy}. However, they focus on the accuracy of the inferred number of users compared to the true number, while our focus is on the sufficiency of that number for making a binary decision which may incur some cost itself. In their setting, data points are also only purchased once, compared to possibly multiple times in our setting.

Many privacy-preserving techniques have been proposed for location privacy such as spatial cloaking~\cite{vergara2011spatialcloaking}, differential privacy~\cite{dwork2014algorithmic, xiao2017loclok} or Geo-Indistinguishablity~\cite{Andres:2013:geo-indis}. These techniques can be applied as the privacy protection mechanism for users' locations in our framework and a pricing model can be derived based on the parameters of these mechanisms, as discussed in Section~\ref{subsec:privacy_pricing}. Therefore, while relevant, these techniques are orthogonal to our work.

\section{Methodology}
\label{sec:methodology}
In this section, we first describe the strategy the buyer would employ to make the open/cancel decision. Given such an open/cancel strategy, we define the \emph{expected incremental profit} ($\eip$) which serves as the basis for our spatial information probing techniques, the SIP and SIP-T algorithms.  

\subsection{The Buyer's Strategy to Make Open/Cancel Decision}
\label{subsec:opening_strategy}
Recall that each buying action $\act^{\buying}(i, \price)$ gives the buyer a noisy data point $\perbpoint_i$.
The noisy data obtained from the buying actions $\seqact_{1:\lasttime-1}$ can help the buyer to make the open/cancel action $\act_{\lasttime}$ as follows.
Assume that after taking actions $\seqact_{1:\lasttime-1}$, the buyer owns a set of noisy data $\perbset$. If there is more than one version of $\perbpoint_i$ from user $i$, the buyer keeps only the most accurate.
Then, from the buyer's perspective, given $\perbpoint_i$, the true location $\dpoint_i$ has the distribution
\begin{equation}
\dpoint_i \sim \mathcal{N}(\perbpoint_i, \sdv_i^2 I)
\end{equation}
The probability $p_i$ of the user $u_i$ being inside $R$ is
\begin{equation} 
p_i = \int_{\region} P(\dpoint_i) d\dpoint_i
\label{eq:probability_inside_region}
\end{equation}
Considering each $p_i$ as the success probability of an independent Bernoulli trial, the probability distribution $P(\nusers | \perbset)$ of the number of users inside region $R$ would be a Poisson binomial distribution~\cite{wang1993poissonbinomialdist} with mean %
as follows:
\begin{align}
	\label{eq:mean_variance_num_users}
	\mu_{\nusers} &= \sum_i p_i %
\end{align}
Given this distribution, the current expected profit for open/cancel actions are $\expectation[\rw(\actvalueopen)] = \uprofit\mu_{\nusers} -\opcost$ and $\expectation[\rw(\actvaluecancel)] = 0$. If the buyer decides to make an open/cancel decision at time $\lasttime$, i.e. $\act_{\lasttime}$ is an open/cancel action $\act^{\opening}$, the best action $\act_{\lasttime}$ would be
\begin{align}
	\label{eq:optimal_action_comparision}
	\act_{\lasttime} = \begin{cases}
		\actvalueopen & \text{if} \quad \uprofit \mu_n  - \opcost > 0  \\
		\actvaluecancel & \text{if} \quad \text{otherwise}
	\end{cases}
\end{align}

Assume that the buyer starts with an empty set of data, which gives $\mu_{\nusers} = 0$. Then, $\expectation[\rw(\actvalueopen)] = -\opcost$. Adding an additional $p_i$ to $\mu_{\nusers}$ in Equation~\ref{eq:mean_variance_num_users} increases  $\mu_\nusers$. Thus by buying more data, the estimate of $\mu_{\nusers}$ increases, thus increasing $\expectation[\rw(\actvalueopen)]$. Therefore, the buyer would want to buy data to increase $\expectation[\rw(\actvalueopen)]$ while also trying to decrease the cost $\sum_{t=1}^{\lasttime-1} \rw(\act_t)$. The buyer would stop buying when they can be confident of making the \emph{open} decision, i.e. $\uprofit \mu_{\nusers}  - \opcost > 0$ or when buying more data, in expectation, would not give more benefit. Thus, $\uprofit \mu_{\nusers}  - \opcost > 0$ can be considered as the \emph{open} trigger for the buyer. We call $\uprofit \mu_{\nusers}  - \opcost > 0$ the \emph{opening condition}.

\subsection{The Expected Incremental Profit (EIP)}
With the strategy to make the open/cancel decision established, we introduce the \emph{expected incremental profit} $\eip(\perbset, \act)$ of taking a buying action $\act$ with the current set $\perbset$ of purchased noisy data. $\eip(\perbset, \act)$ would then serve as the criteria for the buyer to choose a buying action in our algorithms. 

To develop $\eip(\perbset, \act)$, we first calculate the expected profit if the buyer takes $\act^{\opening} = \actvalueopen$ immediately given the current noisy data $\perbset$
\begin{equation}
    \expectation[\rw(\actvalueopen) | \perbset] = \uprofit \sum_i p_i - \opcost
\end{equation}

If the buyer takes a buying action $\act = \act^{\buying}(i, \price)$ and obtains a new noisy data point $\perbpoint'_i$ from the same user with the new probability $p'_i$ of $\dpoint_i \in \region$, replacing the current noisy data $\perbpoint_i$ in $\perbset$ with $\perbpoint'_i$, the expected profit of $\actvalueopen$ is

\begin{equation}
    \expectation[\rw(\actvalueopen) | \perbset, \perbpoint'_i] = \uprofit \big(p'_i + \sum_{j \neq i} p_j \big) - \opcost - \price
\end{equation}
and the incremental profit is:
\begin{align}
    \ip(\perbset, \act, \perbpoint'_i) = \expectation[\rw(\actvalueopen) | \perbset, \perbpoint'_i] - \expectation[\rw(\actvalueopen) | \perbset] \nonumber
        = \uprofit (p'_i - p_i) - \price
\end{align}
If $\perbpoint_i$ does not exist in $\perbset$ (i.e. $\dpoint_i$ was never purchased before), $p_i = 0$.

Since $p'_i$ is unknown before we actually perform the buying action $\act = \act^{\buying}(i, \price)$, we can calculate the expected incremental profit for taking the buying action $\act = \act^{\buying}(i, \price)$ given the current noisy data $\perbset$ as follows:
\begin{align}
    \eip(\perbset, \act) &= \expectation_{\perbpoint'_i} [\ip(\perbset, \act, \perbpoint'_i)] \\
        &= \expectation_{\perbpoint'_i} [\uprofit (p'_i - p_i) - \price] \\
        &= \uprofit \expectation_{\perbpoint'_i} [p'_i |  \perbpoint_i, \sdv_i, \price] - \uprofit p_i - \price \\
        &= \uprofit \int_{\mathbb{R}^2} P(\perbpoint'_i | \perbpoint_i, \sdv_i, \price) p'_i(\perbpoint'_i) d\perbpoint'_i - \uprofit p_i - \price
\end{align}
where $\mathbb{R}$ is the real domain and 
\begin{align}
    p'_i(\perbpoint'_i) &= \int_{\region} P(\dpoint_i) d\dpoint_i \\
    \dpoint_i &\sim \mathcal{N}(\perbpoint'_i, {\sdv'}_i^2 I)
\end{align}

The distribution $P(\perbpoint'_i | \perbpoint_i, \sdv_i, \price)$ of the new noisy point $\perbpoint'_i$ if we takes action $\act = \act^{\buying}(i, \price)$ is the convolution of two distributions: the distribution
$\dpoint_i |\perbpoint_i, \sdv_i \sim \mathcal{N}(\perbpoint_i, \sdv_i^2 I)$
of the true location $\dpoint_i$ given the current noisy data $\perbpoint_i$ and the distribution
$\perbpoint'_i | \dpoint_i, \price \sim \mathcal{N}(\mathbf{0}, {\sdv'}_i^2 I)$
of the next noisy data $\perbpoint'_i$ generated from the true location $\dpoint_i$ where $\sdv'_i$ is the noise magnitude for $\dpoint_i$ at the price $\price$. 
This means the distribution $P(\perbpoint'_i | \perbpoint_i, \sdv_i, \price)$ is
\begin{equation}
    \perbpoint'_i | \perbpoint_i, \sdv_i, \price \sim \mathcal{N}\big(\perbpoint_i, (\sdv_i^2 + {\sdv'}_i^2) I\big)
\end{equation}
The buyer can then choose the next buying action as the action that maximizes the expected incremental profit:
\begin{equation}
    \act^*(\perbset) = \argmax_{\act} \eip(\perbset, \act)
\end{equation}
With $\eip(\perbset, \act)$, we develop two algorithms to help the buyer maximize their profit while maintaining a low purchasing cost.

\subsection{The Spatial Information Probing Algorithms}
We present two greedy algorithms that are based on the \emph{probing} (or \emph{information gathering}) technique. 
The general idea is to utilize the continuity of the price to obtain noisy data at low cost first in order to quickly eliminate uninteresting data before paying a higher price for more accurate data. The two algorithms are called \emph{spatial information probing} (SIP) and SIP with terminals (SIP-T).

\subsubsection{The SIP Algorithm}

The pseudo-code for SIP is shown in Algorithm~\ref{alg:probing_SIP}. SIP has two phases: the first phase is a \emph{pure exploration} phase and the second phase is an \emph{exploration-exploitation} phase. 
In the first phase, the buyer would buy all available data points in a large bounding box around the target region $\region$ at a small \emph{starting price} $\price_0$ (lines~[\ref{alg:line:sip_phase1_start}-\ref{alg:line:sip_phase1_end}]). Then in the second phase (lines~[\ref{alg:line:sip_phase2_start}-\ref{alg:line:sip_phase2_end}]), the buyer repeatedly calculates $\eip$s (lines~[\ref{alg:line:sip_batch_cal_eip_start}-\ref{alg:line:sip_batch_cal_eip_end}]), takes a buying action based on the high potential $\eip$s (line~[\ref{alg:line:sip_buying_using_best_eip_start}, \ref{alg:line:sip_buying_using_best_eip_end}]) and uses the newly purchased noisy data, if any, to guide the next action (line~[\ref{alg:line:sip_cal_next_eip_start}, \ref{alg:line:sip_cal_next_eip_end}]). After each purchase, the buyer also checks the \emph{opening condition} described in Section~\ref{subsec:opening_strategy}. The buyer would stop buying if the best $\eip$ value is negative (line~\ref{alg:line:sip_negative_eip}) which means that in expectation, buying more data would not increase the profit.

\begin{algorithm}[H] 
    
	\caption{The SIP Algorithm} 
	\label{alg:probing_SIP} 
	\begin{algorithmic}[1] 
		\Require Users $\userset$; region $\region$; $\uprofit$; $\opcost$; starting price $\price_0$; price increment factor $\priceincreasefactor$; 
		\Ensure $\actvalueopen$ or $\actvaluecancel$
		
		\State $\perbset \leftarrow \emptyset$; $\profit \leftarrow 0$; $\bestEIPset \leftarrow \{\}$; %
	
        \For{$\dpoint_i \in \userset$} \label{alg:line:sip_phase1_start}
            \State $\perbpoint_i \leftarrow$ Buy $\dpoint_i$ at price $\price_0$ 
            \State $\perbset \leftarrow \perbset \cup \perbpoint_i$; $p_i \leftarrow P(\dpoint_i \in \region | \perbpoint_i)$
            \State $\profit \leftarrow \profit + \uprofit p_i$
            \If{$\profit - \opcost > 0$}
		       \State \Return $\actvalueopen$
		    \EndIf
        \EndFor \label{alg:line:sip_phase1_end}

        \For{$\perbpoint_i \in \perbset$} \label{alg:line:sip_batch_cal_eip_start} \label{alg:line:sip_phase2_start} %
            \State $\act \leftarrow$ $\argmax\limits_{\act} \eip(\perbpoint_i, \act)$ (Algorithm~\ref{alg:cal_best_eip})
            \State $\bestEIPset[\act] \leftarrow \eip(\perbpoint_i, \act)$ \label{alg:line:sip_keep_best_eip_action_1}
        \EndFor \label{alg:line:sip_batch_cal_eip_end}
        
        \While{true} \label{alg:line:sip_start_buying_round}
            \State $\act^{\buying}(i, \price) \leftarrow \argmax_{\act} \bestEIPset[\act]$ \label{alg:line:sip_buying_using_best_eip_start}
            \If{$\bestEIPset[\act^{\buying}(i, \price)] \leq 0$} \label{alg:line:sip_negative_eip}
                \State \Return $\actvaluecancel$
            \EndIf
            \State Delete $\bestEIPset[\act^{\buying}(i, \price)]$ \label{alg:line:sip_cal_next_eip_start}
            
            \State $\perbpoint'_i \leftarrow$ Buy $\dpoint_i$ at price $\price$
            \State $\perbset \leftarrow (\perbset \setminus \perbpoint_i) \cup \perbpoint'_i$ 
            \State $p_i \leftarrow P(\dpoint_i \in \region | \perbpoint_i)$ 
            \State $p'_i \leftarrow P(\dpoint_i \in \region | \perbpoint'_i)$
            \State $\profit \leftarrow \profit - \uprofit p_i + \uprofit p'_i$
            \If{$\profit - \opcost > 0$}
		       \State \Return $\actvalueopen$
		    \EndIf\label{alg:line:sip_buying_using_best_eip_end}

		    \State $\act \leftarrow$ $\argmax\limits_{\act} \eip(\perbpoint_i, \act)$ (Algorithm~\ref{alg:cal_best_eip})
            \State $\bestEIPset[\act] \leftarrow \eip(\perbpoint_i, \act)$ \label{alg:line:sip_keep_best_eip_action_2} \label{alg:line:sip_cal_next_eip_end}
        \EndWhile \label{alg:line:sip_phase2_end}
	\end{algorithmic}
\end{algorithm}

\begin{algorithm}[H] 
	\caption{Find The Best Next Buying Action} 
	\label{alg:cal_best_eip} 
	\begin{algorithmic}[1] 
		\Require $\perbpoint$; $\priveval_i$;  current price $\price$; price increment factor $\priceincreasefactor$; 
		\Ensure Best next buying action $\act^*$
		
		\State $\eipvalue \leftarrow -\infty$
		
		\While{$\price < \priveval_i$}
		    \State $\price \leftarrow \min(\price \times \priceincreasefactor, \priveval_i)$
		    \If{$\eipvalue < \eip(\perbpoint, \act^{\buying}(i, \price))$}
		        \State $\act^* \leftarrow \act^{\buying}(i, \priveval_i)$
		        \State $\eipvalue \leftarrow \eip(\perbpoint, \act^*)$
		    \EndIf
		\EndWhile
		
		\Return $\act^*$
	\end{algorithmic}
\end{algorithm}

Because $\eip(\perbset, \act)$ only depends on $\perbpoint_i$, not any other noisy points, we use $\eip(\perbpoint_i, \act) = \eip(\perbset, \act)$ as $\act^{\buying}(i, \price)$ clearly specifies $\perbpoint_i$. 
Also, to reduce computation time, whenever the buyer calculates $\eip$s, the buyer keeps track of the best next buying action $\act$ for each noisy data point $\perbpoint_i$ and its corresponding value of $\eip$ (line~\ref{alg:line:sip_keep_best_eip_action_1} and~\ref{alg:line:sip_keep_best_eip_action_2}). This best next action is calculated using Algorithm~\ref{alg:cal_best_eip}.

Algorithm~\ref{alg:cal_best_eip} chooses the best next buying action for a noisy data point $\perbpoint_i$ given the most recent purchased price $\price$ and a \emph{price increment factor} $\priceincreasefactor$. 
Even though one can try to calculate the $\eip(\perbpoint_i, \act)$ for all possible prices in $[\price, \priveval_i]$, it would be computationally impractical. 
In addition, the next price which gives the highest $\eip(\perbpoint_i, \act)$ may be too close to $\price$ to be informative enough (i.e. significant change in $p_i$) to make good progress. 
To overcome these obstacles, Algorithm~\ref{alg:cal_best_eip} operates in multiple rounds and increases the potential price in each round by a factor $\priceincreasefactor$ until the potential price is at least $\priveval_i$. The factor $\priceincreasefactor$ allows the next purchase for $\dpoint_i$ to be significantly more informative than the current $\perbpoint_i$, besides reducing computation. 

\subsubsection{The SIP-T Algorithm}
One issue with SIP is that, 
for $\dpoint_i$ that is actually inside $\region$ but closer to the edges, the first low-price purchases may give unexpectedly low values for the probability of $\dpoint_i$ being inside $\region$, making the increment in the expected profit become smaller than the next price. 
This issue may cause the highest value of $\eip(\perbpoint_i, \act)$ to be negative, thus, SIP may stop buying earlier than expected. We propose the SIP-T algorithm to address this issue. 

SIP-T defines a \emph{terminal} belief (or terminal) $\terminal_i$ for each data point $\dpoint_i$ and only stops buying when either all data points are in their terminals or the \emph{opening condition} is satisfied. The terminal $\terminal_i$ specifies that the buyer can be certain about whether $\dpoint_i$ is inside $\region$ or not. The \emph{terminating} condition determines if $\perbpoint_i$ is at $\terminal_i$.

Although a threshold for $p_i$ can be used for the terminating condition (e.g. $p_i < 0.05$), the high-magnitude noise when purchasing data at low prices make this approach ineffective. Instead, SIP-T uses the standard deviation $\sdv_i$ of the noise to check for the terminating condition. More specifically, for a noisy data point $\perbpoint_i$ with the standard deviation $\sdv_i$ of the noise, $\perbpoint_i$ is at $\terminal_i$ if in each dimension $\perbpoint_i$ is either (1) inside $\region$ with the distance to each edge at least $\sigmafactor \sdv_i$ or (2) outside $\region$ with the distance to the closest edge at least $\sigmafactor \sdv_i$. We arbitrarily choose $\sigmafactor = 2$.

With the terminal condition defined as above, SIP-T is SIP with two changes. 
The first change is that, in line~\ref{alg:line:sip_negative_eip}, it returns $\actvaluecancel$ when $\bestEIPset = \emptyset$ instead of $\bestEIPset[\act^{\buying}(i, \price)] \leq 0$. 
The second change is that after buying a data point and updating the expected profit in line~\ref{alg:line:sip_buying_using_best_eip_end}, it checks for the terminating condition, and if the condition holds, it immediately continues to the next purchase round (line~\ref{alg:line:sip_start_buying_round}) instead of calculating next best buying action for that data point.

\section{Experimental Evaluation}
\label{sec:experiments}
We experimentally evaluate our probing algorithms on a real-world dataset with various settings for users' data, the buyer's decisions and algorithmic parameters. 

\label{subsec:exp_setup}
\subsection{Datasets}
We experiment on the Gowalla dataset from the SNAP project~\footnote{https://snap.stanford.edu/data/loc-gowalla.html} with 6,442,890 check-ins of 196,591 users over the period of February 2009 to October 2010. Each check-in includes a user id, check-in time, latitude and longitude.

We collect check-ins within a Los Angeles boundary defined by a bounding box from a southwest corner at (-118.684687, 33.699675) to northeast corner (-118.144458, 34.342324) in degrees of latitude and longitude. We then keep only one random check-in per user, which results in a subset of 5827 check-ins. 

\begin{figure}[htbp!]
\centering
\includegraphics[width=.5\linewidth]{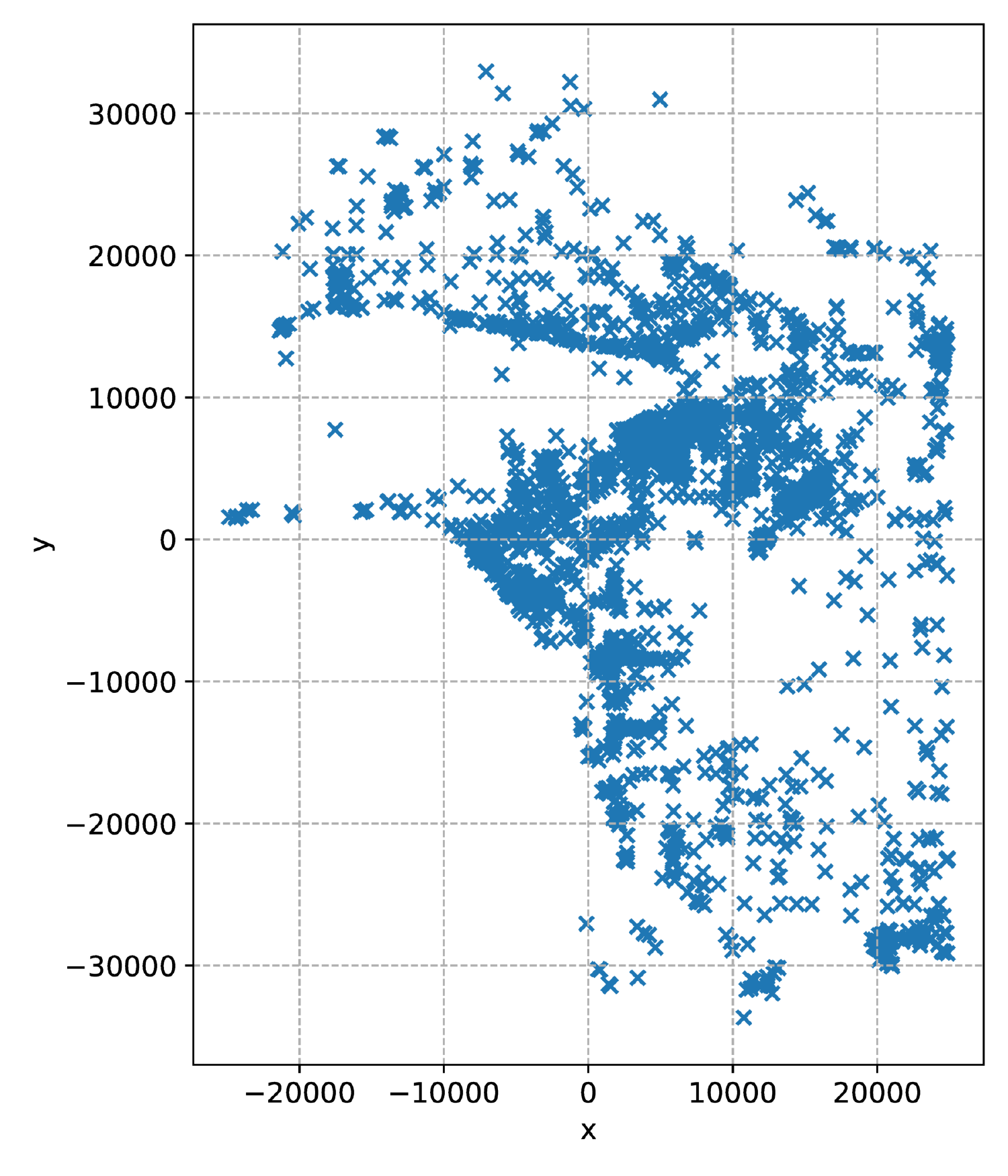}
\caption{Check-ins of Gowalla users in Los Angeles converted to local Euclidean coordinates, 1 check-in per user.} 
\label{fig:checkins_LA_1_per_user}
\end{figure}

The radian (lat, long) coordinates are then converted to a locally planar Cartesian coordinate system with the mid-point of the latitude/longitude bounding box as the reference point.
The bounding box in local Euclidean coordinates $(x, y)$ is from the southwest corner at (-25,000, -35,000) to the the northeast corner at (25,000, 35,000) in meters. Figure~\ref{fig:checkins_LA_1_per_user} shows the check-ins in Los Angeles area within the bounding box.




\subsection{Evaluation Metrics}
To avoid any bias towards any position of the target region $\region$, we evaluate performance of each algorithm over a collection of regions $\regionset$. Our collection of regions is a grid over the bounding box. Each cell of the grid is considered as a target region for making an open/cancel decision. The evaluation metrics includes the \emph{Average Realized Profit} (ARP), the \emph{Median Realized Profit} (MRP) and the \emph{Recall} as explained in the following.

\textbf{The ARP and MRP metrics.} For each target region $\region \in \regionset$, the net profit of the buyer's action sequence $\seqact_{\region}$ is $\rw(\seqact_{\region})$, calculated using Equation~\ref{eq:max_expected_profit_repeated}.
Then ARP and MRP are defined as the mean and median values of $\{\rw(\seqact_{\region}) | \region \in \regionset\}$. 
The reason we use MRP, in addition to ARP, is because there are often a few popular regions that have many users and become outliers in term of ARP since opening a restaurant there can yield extremely high profit compared to other regions. MRP is more robust to the effect of outliers and, as discussed later, can better reflect the cost spent by each algorithm.

\textbf{Recall.} Given the ground truth data, for each target region $\region$, if the opening condition holds/does not hold for $\region$, $\region$ is considered as a \emph{positive/negative} region. Similarly, the value $\actvalueopen/\actvaluecancel$ of the open/cancel action $\act_{\region}^{\opening}$ is considered as \emph{positive/negative} decision. Consequently, recall can be calculated for the set $\{\act_{\region}^{\opening} | \region \in \regionset\}$ as the ratio of the positive regions the algorithm can find.

\subsection{Baselines}
We compare our algorithms to several baselines. As mentioned in Section~\ref{sec:background_related_work}, since standard POMDP algorithms are computationally infeasible for our problem, we do not consider them as baselines.

The first baseline is the \emph{Oracle} algorithm which simply knows the most accurate location data of all users. Oracle is used as a benchmark to show the maximum value of each metric that any algorithm can achieve, even though this is unlikely since an algorithm usually needs to purchase data in order to make an open/cancel decision and the price of purchasing data may surpass the profit.

The second baseline is the utility maximization algorithm in~\cite{singla2018price}, referred here as the \emph{PoI algorithm}. 
With PoI, each point $\dpoint_i$ is assigned a value (called a ``grade"), which is the extra expected profit $\poiextravalue$ such that the expected profit the buyer may earn by buying $\dpoint_i$ at a certain price to obtain new noisy data is the same as the expected profit from stopping buying $\dpoint_i$ and receiving $\poiextravalue$.
The grade $\poiextravalue$ is calculated based on the price of $\dpoint_i$, the profit per user $\uprofit$ and a uniform probability of $\dpoint_i$ being inside the region $\region$. 
PoI then ranks data points based on their grades and then repeatedly buys data points \emph{at full price} until the maximum grade is non-positive or the opening condition holds.

The other baseline is the \emph{Fixed Maximum Cost} (FMC) algorithm that spends a fixed amount to buy all data points at the same price. As mentioned, the challenge with the FMC is how to predict the fixed amount. We used $0.1\%$, $1\%$ and $2\%$ of the fixed cost $\opcost$ to derive three variations, called \emph{FMC-0.1, FMC-1} and \emph{FMC-2}, respectively.

\subsection{Parameter Setup}
\label{subsec:param_setup}
The gross margin per user $\uprofit$ is set to $\uprofit \in \{50, \mathbf{100}, 150, 200\}$ in US dollars. The values are chosen based on the annual gross margin of some fast food chains such as Starbucks, McDonald's or Del Taco, divided by their total number of customers, which can be found in their annual reports. The bold value indicates the default value used in the experiments where this parameter is fixed. 

Instead of fixing values for the fixed cost $\opcost$, which is difficult to know in the real world, we consider the ratio $\nusers_0 = \opcost/\uprofit$ which can be seen as the minimum number of users that the buyer would need to $\actvalueopen$, because if $\nusers > \nusers_0$, then $\uprofit \nusers > \opcost$, which means the opening condition holds. 
We call this ratio the \emph{minimum user threshold} and set it to $\nusers_0 \in \{200, 300, \mathbf{400}, 500, 600\}$. These values are derived from~\cite{gase2019understanding_opening_threshold} where the authors showed that the average number of restaurants was 2.3 (SD, 1.8) per 1,000 residents in Los Angeles County, yielding about 435 residents per restaurant. When $\uprofit$ is fixed at $100$, these values yield $\opcost \in \{\text{20,000}, \text{30,000}, \text{40,000}, \text{50,000}, \text{60,000}\}$. 

For users' privacy valuation, we simulate the privacy level and sensitivity values from two uniform distributions $\privlvl_i \sim \mathit{Unif}(0, \privscale)$ and $\sensitivity_{i, j} \sim \mathit{Unif}(0, \privscale)$ then multiply them to get $\priveval_i = \privlvl_i \sensitivity_{i, j}$. These distributions mean that each user has general privacy levels from $0$ to $\privscale$, and their own data points have sensitivity levels from $0$ to $\privscale$. The scale $\privscale$ is set to $\privscale \in \{1, 2, \mathbf{3}, 4, 5\}$.

For the size of the regions \region, we experimented with different $\regionsize \times \regionsize$ region sizes with $\regionsize \in \{\text{2,500}, \textbf{5,000}, \text{7,500}, \text{10,000}\}$ meters, which creates from 35 to 560 regions in the test grid. These values are derived from the study~\cite{melaniphy2007restaurantselection}. 

For the parameters of SIP and SIP-T, the starting price $\price_0$ is set to $\price_0 \in \{0.0001, \mathbf{0.001}, 0.002, 0.003, 0.004\}$, and the price increment factor $\priceincreasefactor$ is set to $\priceincreasefactor \in \{1.5, \mathbf{2}, 3.5, 5\}$ to reflect small and large increment factors. The scaling factor $\ratesdv$ is fixed at $\ratesdv = 20$, which would render the standard deviation of noise to be $\{20, 40, 80, \dots\}$ for the ratio $\frac{\priveval_i}{q} = \{1, 0.5, 0.25, \dots\}$. These values are close to the noise magnitude of real-world location noise such as GPS, Wi-Fi or cell towers. Finally, the experiments for each region is executed in a single core of an Intel$^{\tiny{\text{\textregistered}}}$ Core\textsuperscript{TM} i9-9980XE CPU. %

\begin{figure}[t!]
\centering
  \subfigure[ARP]{\includegraphics[scale=0.22]{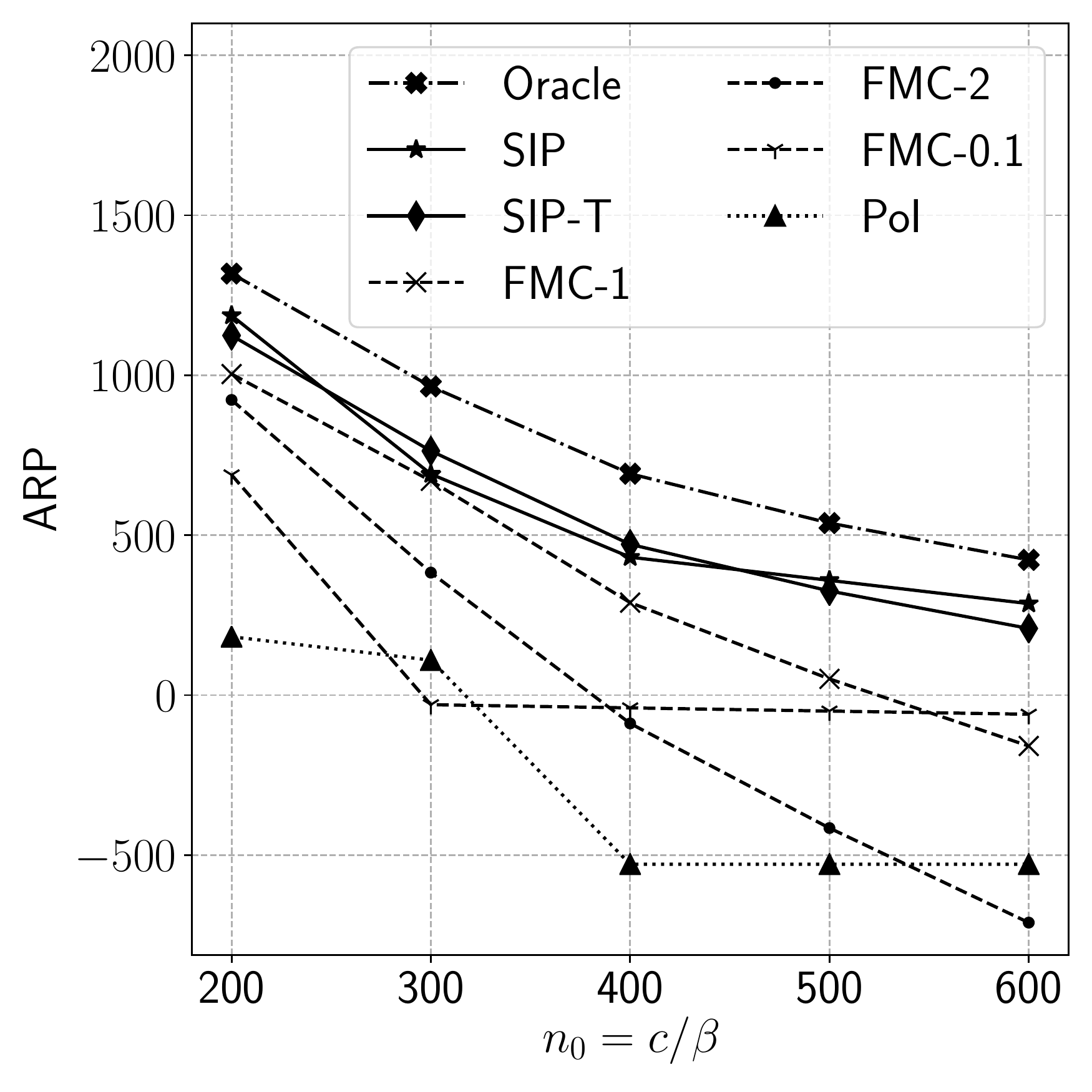}\label{fig:exp_vary_opening_ratio_linear_avg_adjusted_payoff}}%
  \subfigure[MRP]{\includegraphics[scale=0.22]{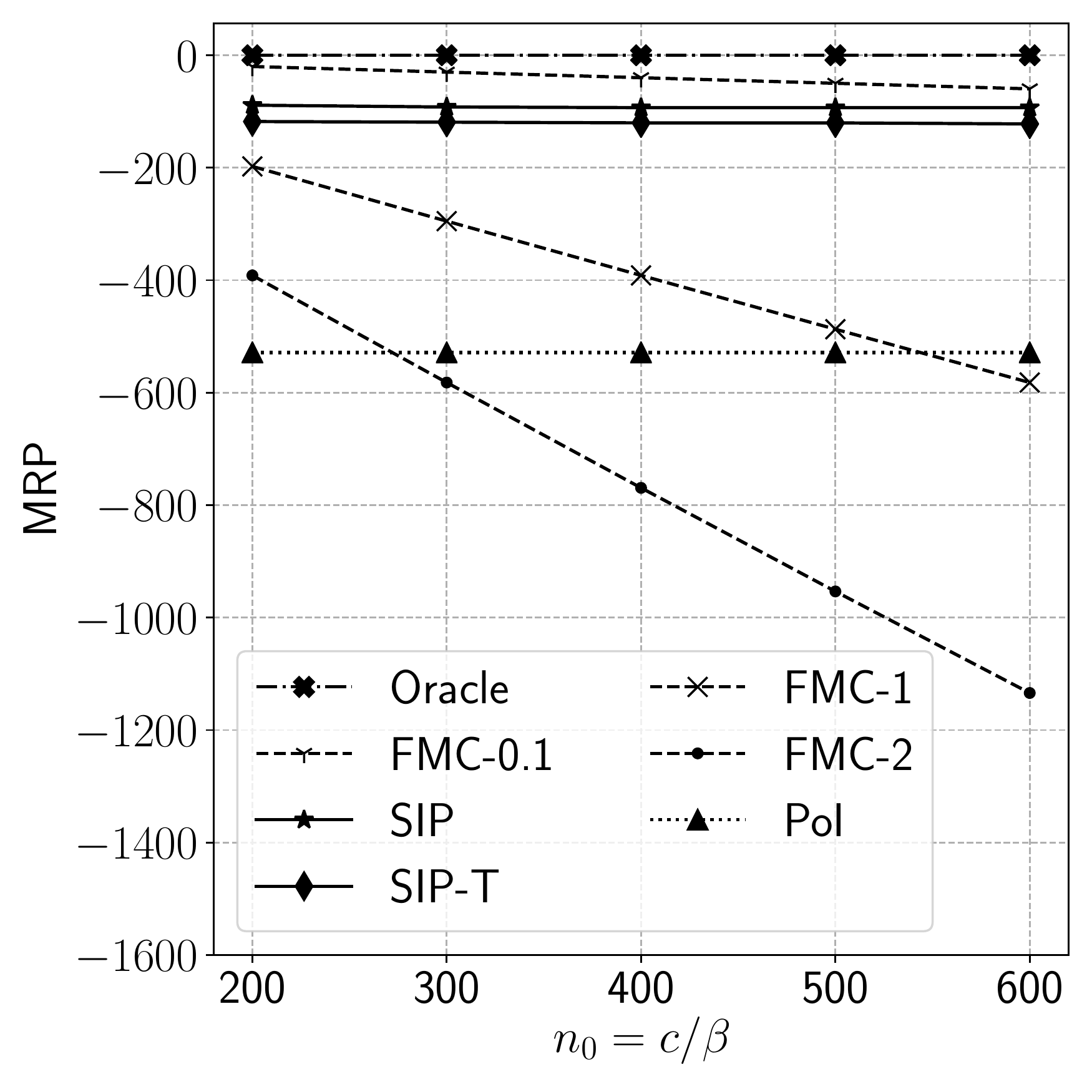}\label{fig:exp_vary_opening_ratio_linear_median_adjusted_payoff}}%
  
  \subfigure[Recall]{\scalebox{0.75}{
  \bgroup
        \def\arraystretch{1}%
        \addvbuffer[0pt 0pt]{\begin{tabular}[b]{l||r|r|r|r|r|}
         & \multicolumn{1}{c|}{\textbf{200}} & \multicolumn{1}{c|}{\textbf{300}} & \multicolumn{1}{c|}{\textbf{400}} & \multicolumn{1}{c|}{\textbf{500}} & \multicolumn{1}{c|}{\textbf{600}} \\ \hline \hline
        \textbf{FMC-0.1} & 0.17 & 0 & 0 & 0 & 0 \\ \hline
        \textbf{FMC-1} & 0.83 & 1 & 0.67 & 1 & 1 \\ \hline
        \textbf{FMC-2} & 0.83 & 1 & 0.67 & 1 & 1 \\ \hline
        \textbf{PoI} & 0.17 & 0.25 & 0 & 0 & 0 \\ \hline
        \textbf{SIP} & 0.83 & 0.5 & 0.33 & 0.5 & 1 \\ \hline
        \textbf{SIP-T} & 0.83 & 1 & 0.67 & 1 & 1 \\ \hline 
        \end{tabular}\label{tbl:vary_opening_ratio_recall}}
    \egroup}}
  \caption{The effect of the minimum user threshold $\nusers_0$}
  \label{fig:exp_vary_opening_ratio}
\end{figure}

\subsection{Experimental Results}  %

\subsection{The effect of profit model parameters}

\subsubsection{The minimum customer threshold $\nusers_0$}

We first show in Figure~\ref{fig:exp_vary_opening_ratio} the effect of the minimum customer threshold $\nusers_0$. 
SIP and SIP-T consistently outperform other algorithms in most cases.

The average realized profit (ARP) decreases when $\nusers_0$ increases, since there are fewer regions that have enough users to open a restaurant (Figure~\ref{fig:exp_vary_opening_ratio_linear_avg_adjusted_payoff}). However, while the ARP of other algorithms sharply decreases, SIP and SIP-T maintain their performance compared to Oracle. Note that Oracle is assumed to know the accurate data of all users. The actual cost to obtain such accurate data is about \textit{1.8 million} USD. Besides, although ARP of \emph{FMC-1} is sometimes similar to that of SIP and SIP-T, we emphasize that FMC comes with no principles for deciding how much to spend. Therefore, in some scenarios, it just got lucky. For example, FMC can easily suffer from underspending (such as \emph{FMC-0.1}) or overspending (such as \emph{FMC-2}). 

The median realized profit (MRP) further demonstrates the superiority of SIP and SIP-T compared to other algorithms (Figure~\ref{fig:exp_vary_opening_ratio_linear_median_adjusted_payoff}). Since the decision for the majority of the regions in the grid should be $\actvaluecancel$, Oracle has a zero MRP and other algorithms have negative MRPs, which are the median amount they spent on purchasing data. MRP values of SIP and SIP-T are stable and several times higher than those of other algorithms, except for \emph{FMC-0.1} which was underspending. 

\begin{figure}[htpb!]
  \centering
  \subfigure[Amount spent per region]{\includegraphics[scale=0.22]{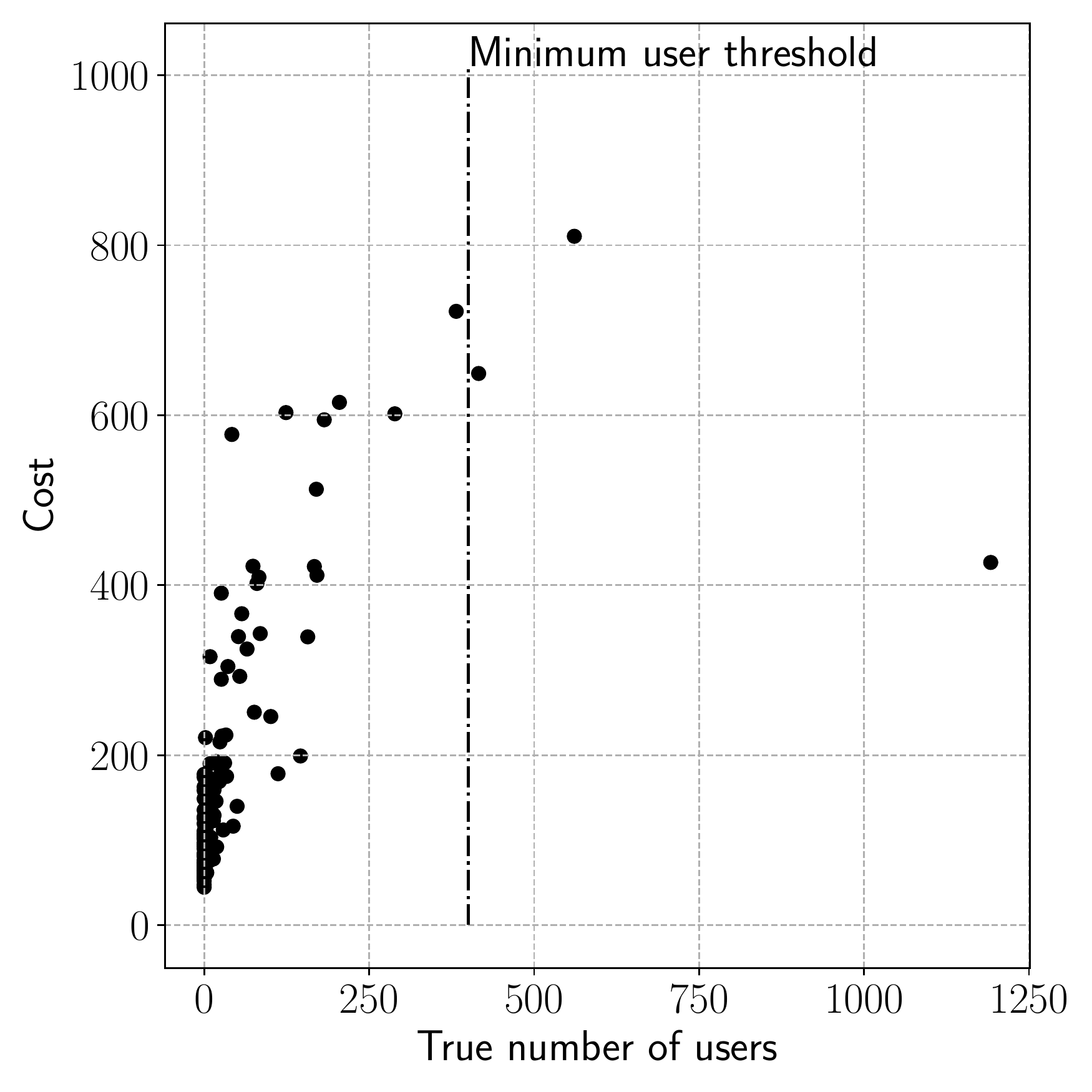}\label{fig:exp_true_count_vs_cost_random_priv_False}}%
  \subfigure[Amount spent per data point]{\includegraphics[scale=0.22]{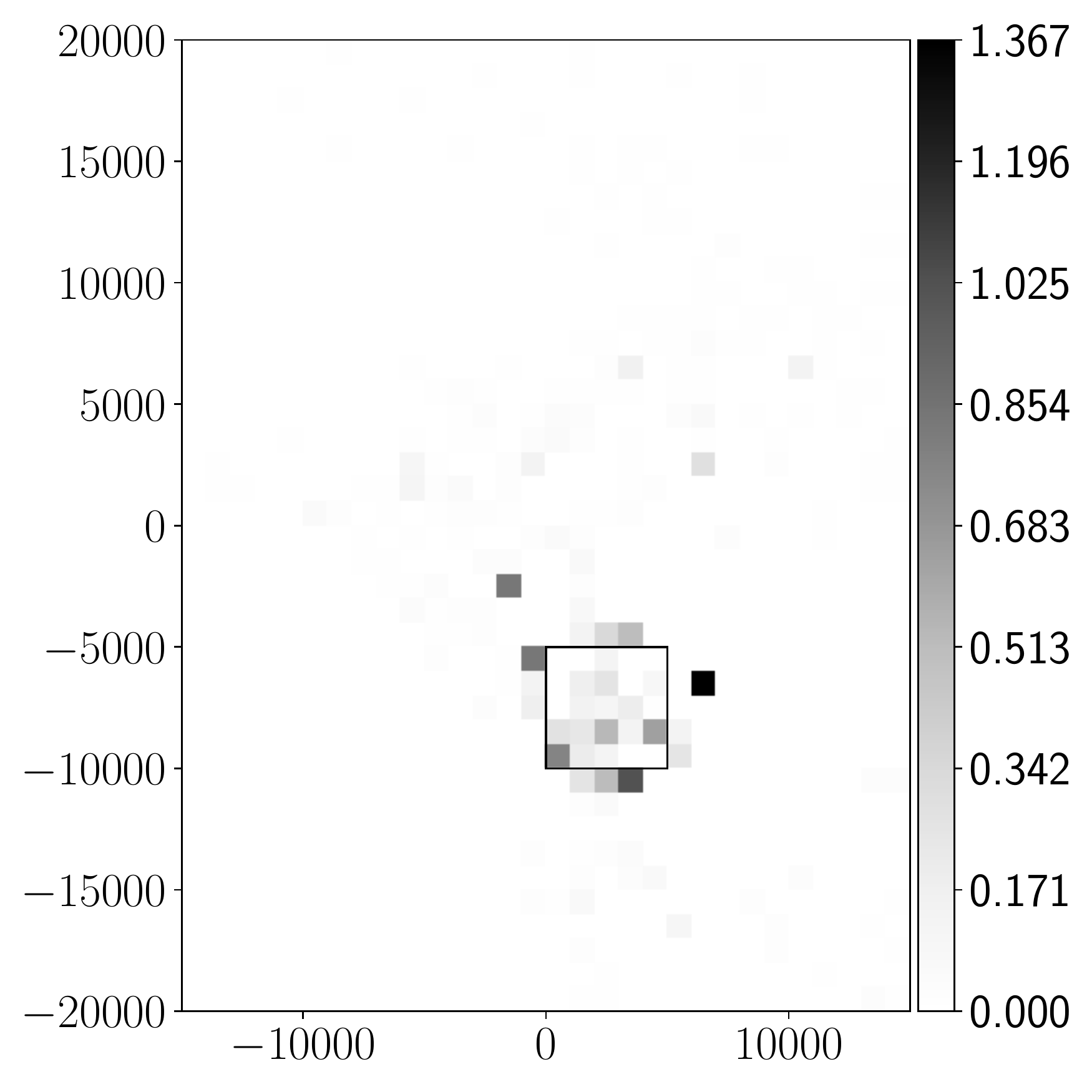}\label{fig:exp_heatmap_random_priv_False}}
  \caption{Illustrations for the amount spent by the SIP}
  \label{fig:exp_sip_illustrations}
\end{figure}

While often spending much less than the other algorithms, SIP and SIP-T can still make correct $\actvalueopen$ decisions for the regions where the buyer should decide $\actvalueopen$. This is shown in Table~\ref{tbl:vary_opening_ratio_recall} where recall of SIP and SIP-T is comparable to \emph{FMC-1} and \emph{FMC-2} which spent significantly more.

The reason for the superior performance of SIP and SIP-T are their highly adaptive nature: adaptive to the number of users in the target region and to the position of each data point relative to the target region. Figure~\ref{fig:exp_sip_illustrations} illustrates the amount SIP spent for each region against the true number of users (Figure~\ref{fig:exp_true_count_vs_cost_random_priv_False}) and for each data point (Figure~\ref{fig:exp_heatmap_random_priv_False}). The amount spent tends to be higher for regions that have the true number of users closer to the minimum user threshold and higher for data points closer to the edges of a target region. This is because the true state being inside or outside a region is harder to identify for users closer to the edge, thus requiring more accurate data, which costs more. Although PoI dynamically decides which data point to buy, it does not perform well compared to SIP and SIP-T because it bought data at full price.

While SIP and SIP-T have comparable ARP, SIP has a higher MRP but a lower recall than that of SIP-T. This is because SIP-T continues to buy a data point at higher accuracy even though the EIP of the data point can be negative. Hence SIP-T would spend more than SIP but have more accurate location information.

The general trade-off is that spending more money buys more accurate data, which helps make more accurate open/cancel decisions. However, spending excessively may decrease the final profit, because the high cost of purchasing data may surpass the profit. SIP and SIP-T are two alternatives that balance this trade-off with different foci: on the expected profit per data point vs. distinguishing whether a data point is inside or outside the target region. 

\begin{figure}[htpb!]
  \centering
  \subfigure[ARP]{\includegraphics[scale=0.22]{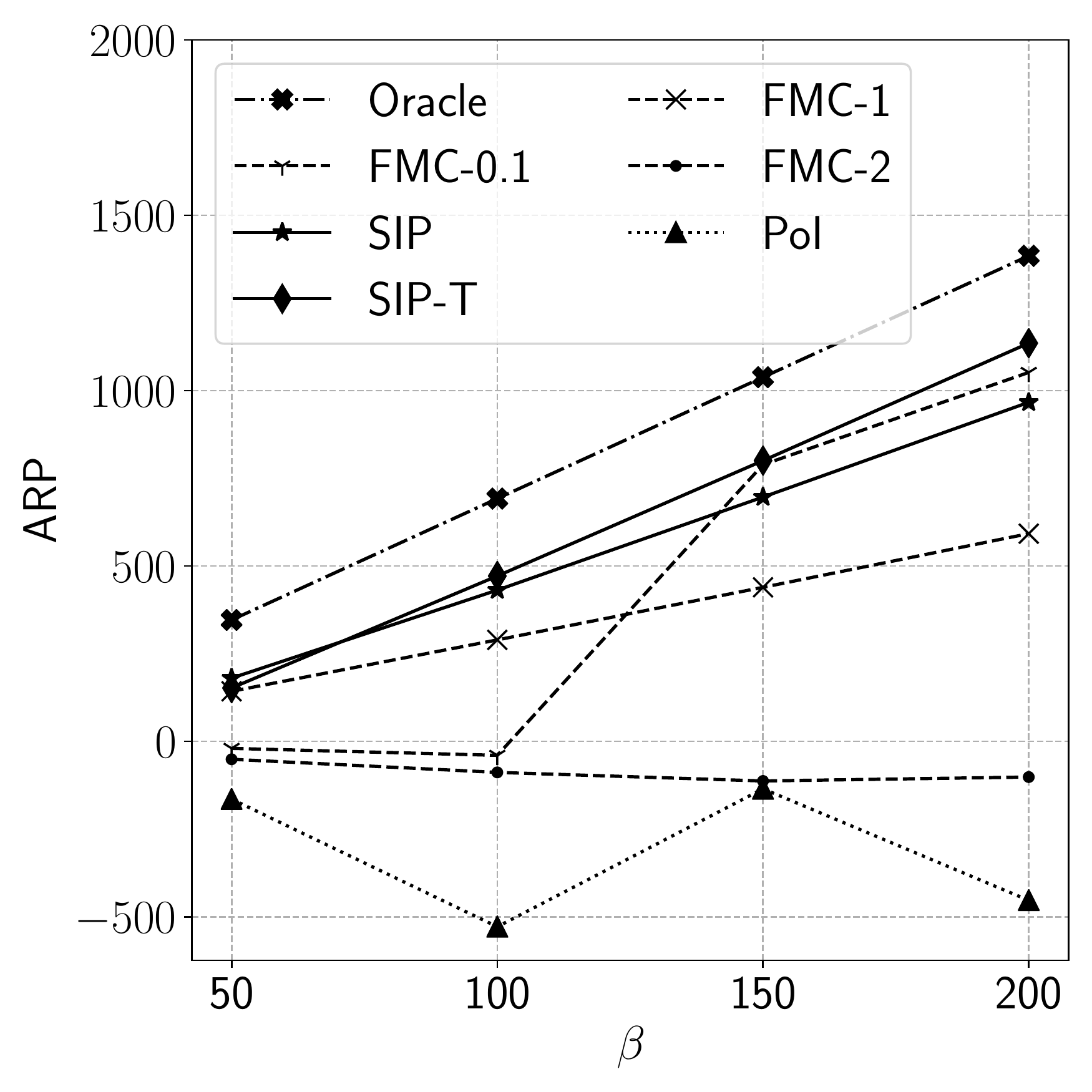}\label{fig:exp_vary_profit_per_user_linear_avg_adjusted_payoff}}%
  \subfigure[MRP]{\includegraphics[scale=0.22]{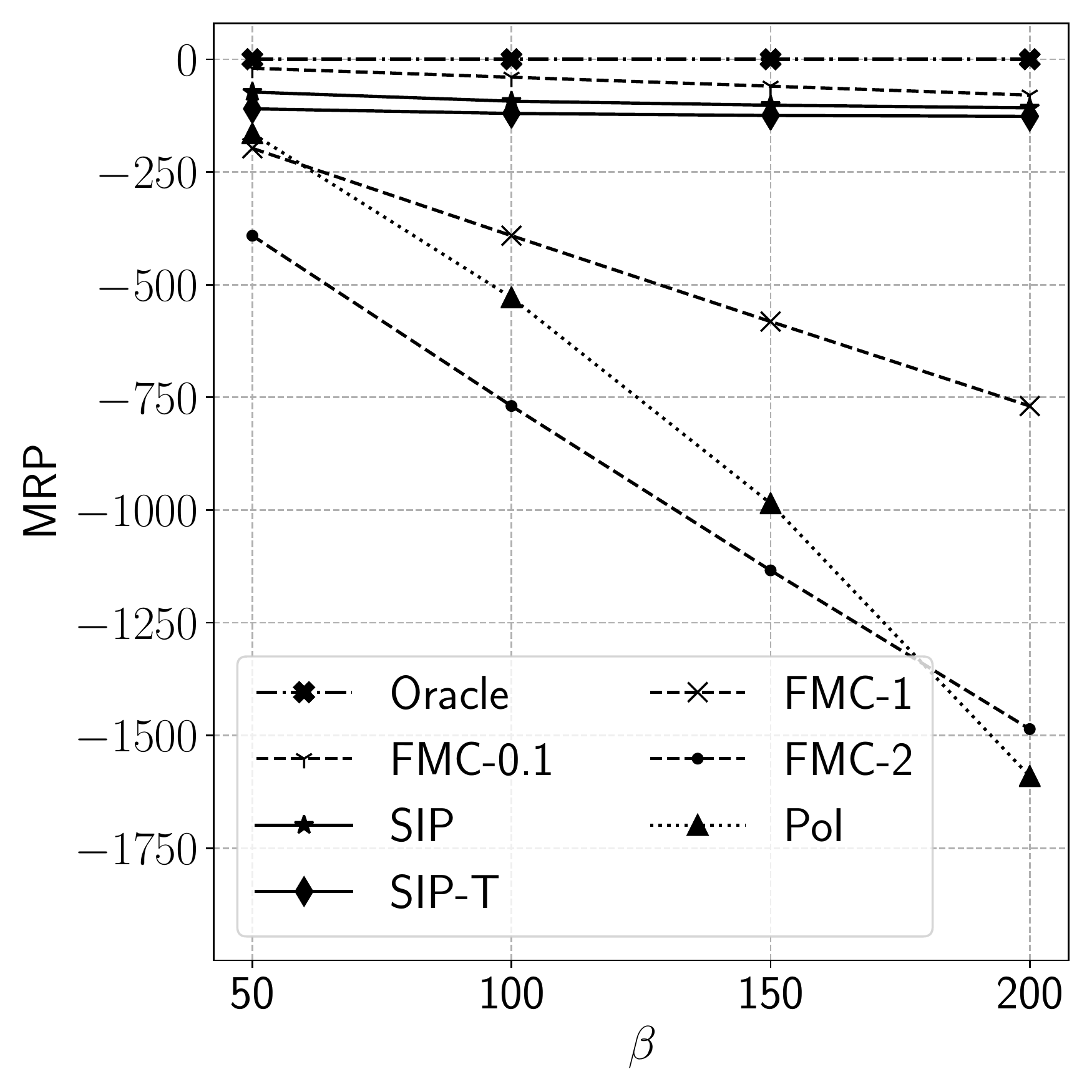}\label{fig:exp_vary_profit_per_user_linear_median_adjusted_payoff}}%
  
  \subfigure[Recall]{\scalebox{0.75}{
  \bgroup
        \def\arraystretch{1}%
        \addvbuffer[0pt 0pt]{\begin{tabular}[b]{l||r|r|r|r|}
     & \multicolumn{1}{c|}{\textbf{50}} & \multicolumn{1}{c|}{\textbf{100}} & \multicolumn{1}{c|}{\textbf{150}} & \multicolumn{1}{c|}{\textbf{200}} \\ \hline \hline
    \textbf{FMC-0.1} & 0 & 0 & 0.33 & 0.33 \\ \hline
    \textbf{FMC-1} & 0.67 & 0.67 & 0.67 & 0.67 \\ \hline
    \textbf{FMC-2} & 0.67 & 0.67 & 0.67 & 1 \\ \hline
    \textbf{PoI} & 0 & 0 & 0.33 & 0.33 \\ \hline
    \textbf{SIP} & 0.33 & 0.33 & 0.33 & 0.33 \\ \hline
    \textbf{SIP-T} & 0.67 & 0.67 & 0.67 & 0.67 \\ \hline 
    \end{tabular}\label{tbl:exp_vary_profit_per_user_recall}}
    \egroup}}
  \caption{The effect of the gross margin per user $\uprofit$}
  \label{fig:exp_vary_profit_per_user}
\end{figure}

\subsubsection{The profit per user $\uprofit$}

The effect of the gross margin (or gross profit) per user $\uprofit$ is shown in Figure~\ref{fig:exp_vary_profit_per_user}. As $\uprofit$ increases, the buyer can gain more per user, thus, gaining higher ARP. However, as minimum user threshold $\nusers_0$ is fixed, when $\uprofit$ increases, $\opcost$ also increases, thus, increasing the amount the FMC-based algorithm spends. That is why MRP decreases for the FMC-based algorithms. 

With a higher value of $\uprofit$, the EIP of a data point at a price also increases. Increasing EIP leads to a higher chance that such data would be purchased at a higher price. This explains the slight decrease in MRP of SIP. Since SIP-T does not rely on EIP to stop buying a data point, its MRP does not change. Recall of SIP and SIP-T remain comparable to \emph{FPC-1} and \emph{FMC-2} while spent significant less. The result from the minimum customer threshold $\nusers_0$ and the profit per user $\uprofit$ show that SIP and SIP-T can give consistent and high results for different profit model parameters.

\subsection{The effect of query parameters and user's data}

\subsubsection{The target region $\regionsize$}
\begin{figure}[htpb!]
  \centering
  \subfigure[ARP]{\includegraphics[scale=0.22]{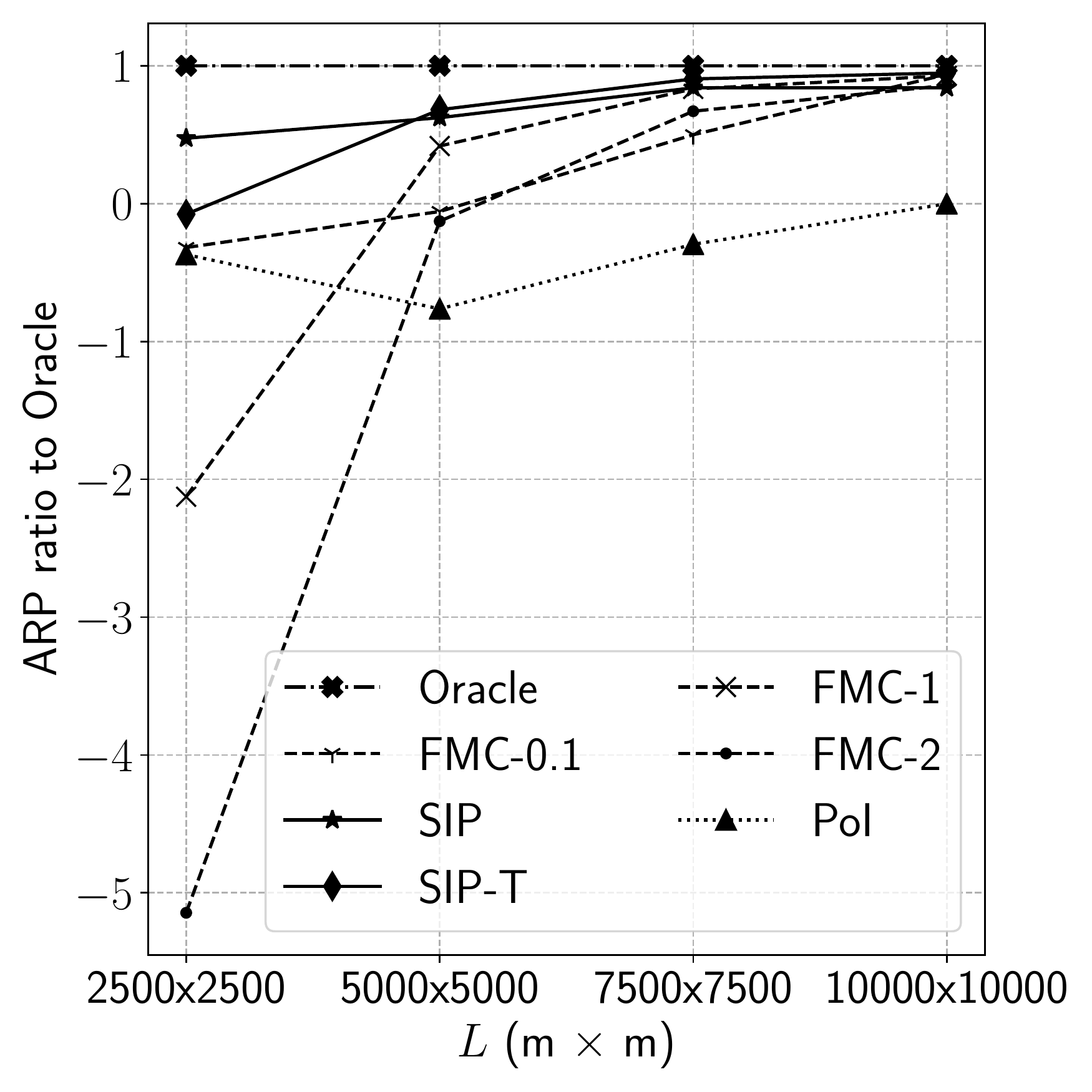}\label{fig:exp_vary_grid_cell_len_linear_avg_adjusted_payoff}}%
  \subfigure[MRP]{\includegraphics[scale=0.22]{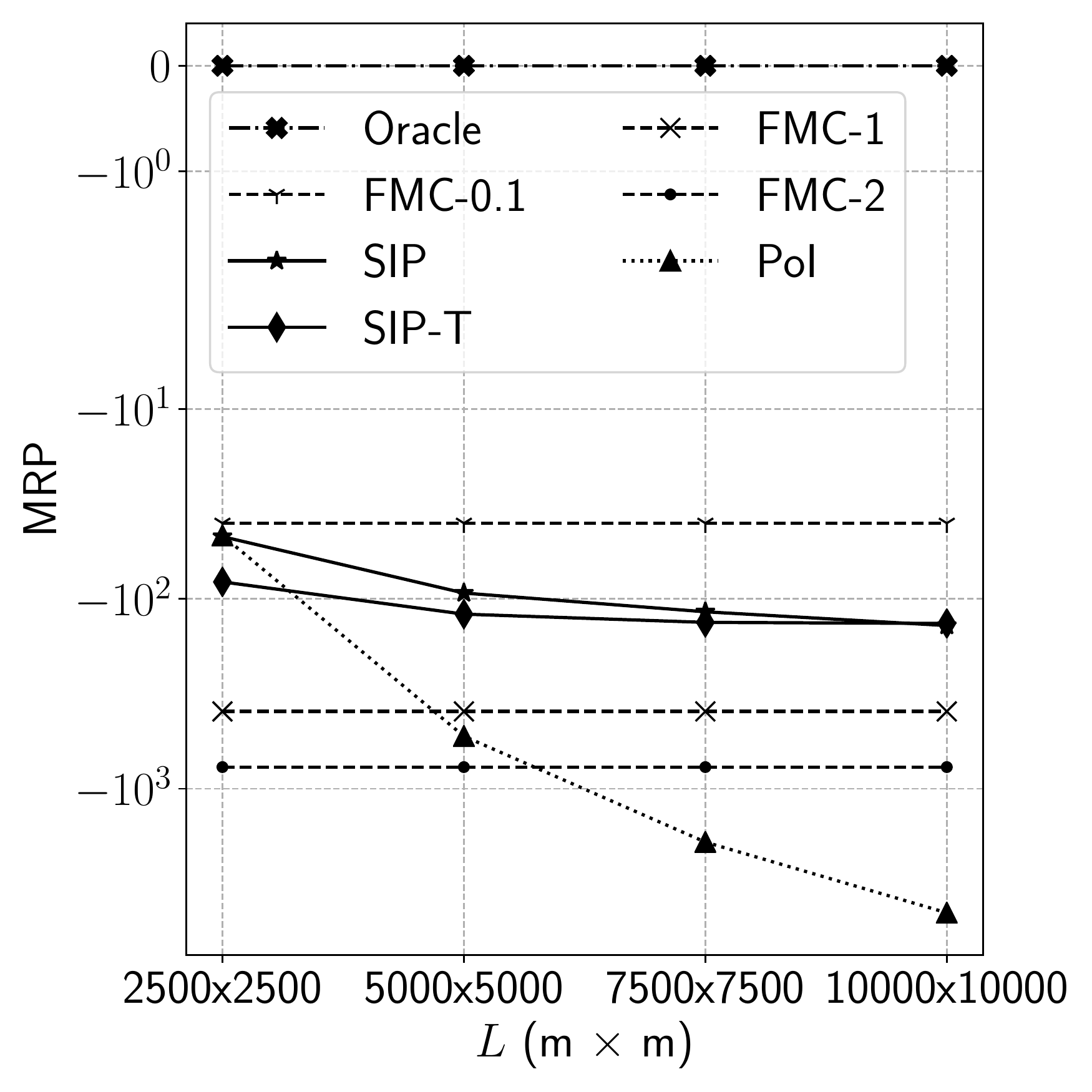}\label{fig:exp_vary_grid_cell_len_linear_median_adjusted_payoff}}%
  
  \subfigure[Recall]{\scalebox{0.75}{
  \bgroup
        \def\arraystretch{1}%
        \addvbuffer[0pt 0pt]{\begin{tabular}[b]{l||r|r|r|r|}
         & \multicolumn{1}{c|}{\textbf{2500}} & \multicolumn{1}{c|}{\textbf{5000}} & \multicolumn{1}{c|}{\textbf{7500}} & \multicolumn{1}{c|}{\textbf{10000}} \\ \hline \hline
        \textbf{FMC-0.1} & 0 & 0 & 0.2 & 0.8 \\ \hline
        \textbf{FMC-1} & 1 & 0.67 & 1 & 1 \\ \hline
        \textbf{FMC-2} & 1 & 0.67 & 1 & 1 \\ \hline
        \textbf{PoI} & 0 & 0 & 0.2 & 0.4 \\ \hline
        \textbf{SIP} & 1 & 0.33 & 0.6 & 0.6 \\ \hline
        \textbf{SIP-T} & 1 & 0.67 & 1 & 1 \\ \hline 
        \end{tabular}\label{tbl:exp_vary_grid_cell_len_recall}}
    \egroup}}
  \caption{The effect of the size $\regionsize$ of the target region}
  \label{fig:exp_vary_grid_cell_len}
\end{figure}

Figure~\ref{fig:exp_vary_grid_cell_len} shows the effect of the size $\regionsize$ of the target region. Because the difference between values of ARP and MRP for different values of the size $\regionsize$ is large, ARP is shown as the ratio to ARP of Oracle and MRP is shown on a logarithmic scale. With a larger region (i.e. a higher value of $\regionsize$), the buyer can gain more users, resulting in a higher profit in general. When the region size is very large (e.g. $\regionsize = \text{10,000}$), all algorithms achieve comparable ARP. However, when $\regionsize$ becomes smaller, SIP and SIP-T can maintain good performance while other methods suffer. Also, when $\regionsize$ gets smaller, with a uniform probability, the probability of a data point being inside a region also becomes smaller. This makes PoI exclude many data points from buying, which may eventually not buy any data points and have a $0$ recall. SIP and SIP-T, again, can maintain good ARP and recall while spent significantly less. 

\subsubsection{The scale $\privscale$ of users' privacy distributions}

\begin{figure}[htpb!]
  \centering
  \subfigure[ARP]{\includegraphics[scale=0.22]{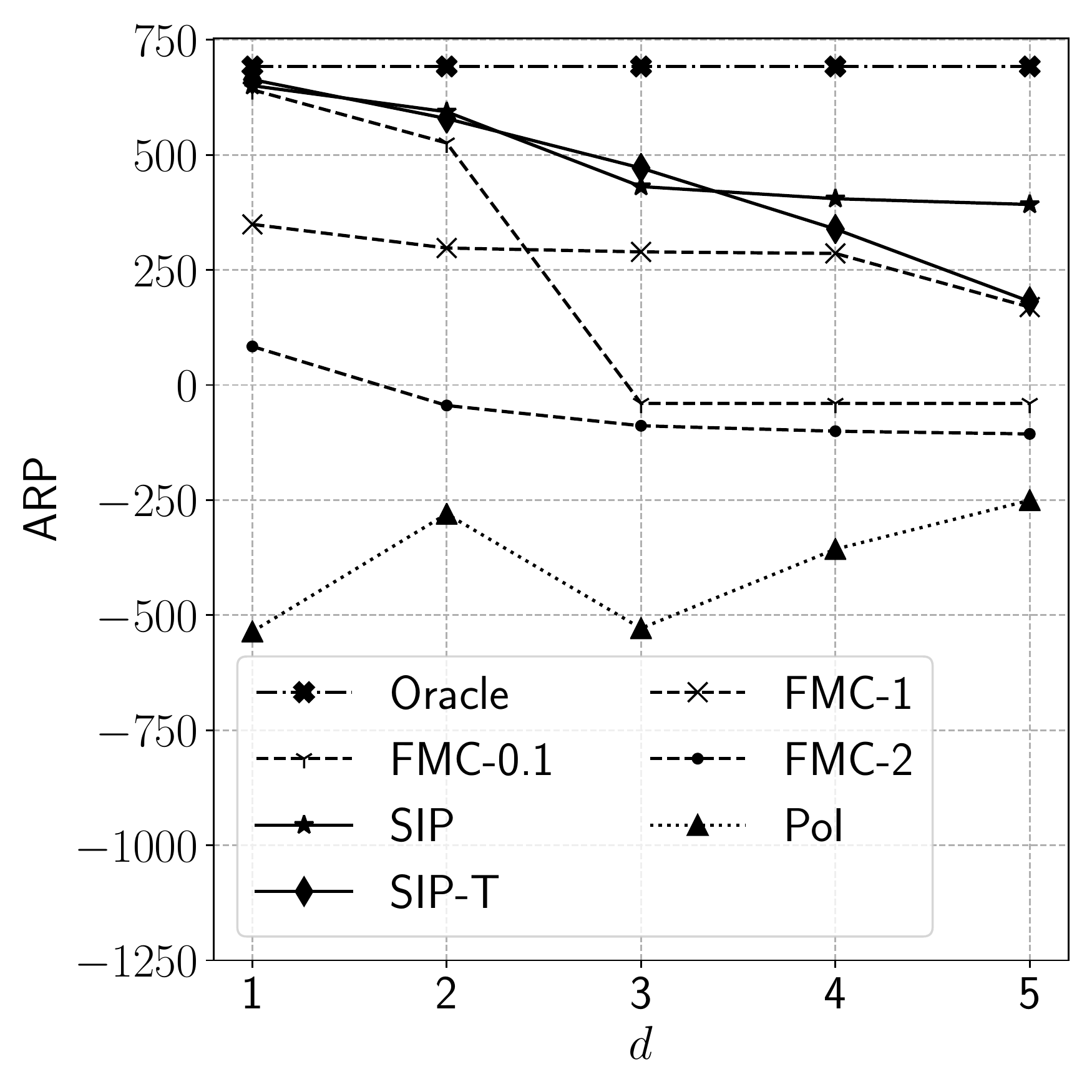}\label{fig:exp_vary_privacy_valuation_scale_linear_avg_adjusted_payoff}}%
  \subfigure[MRP]{\includegraphics[scale=0.22]{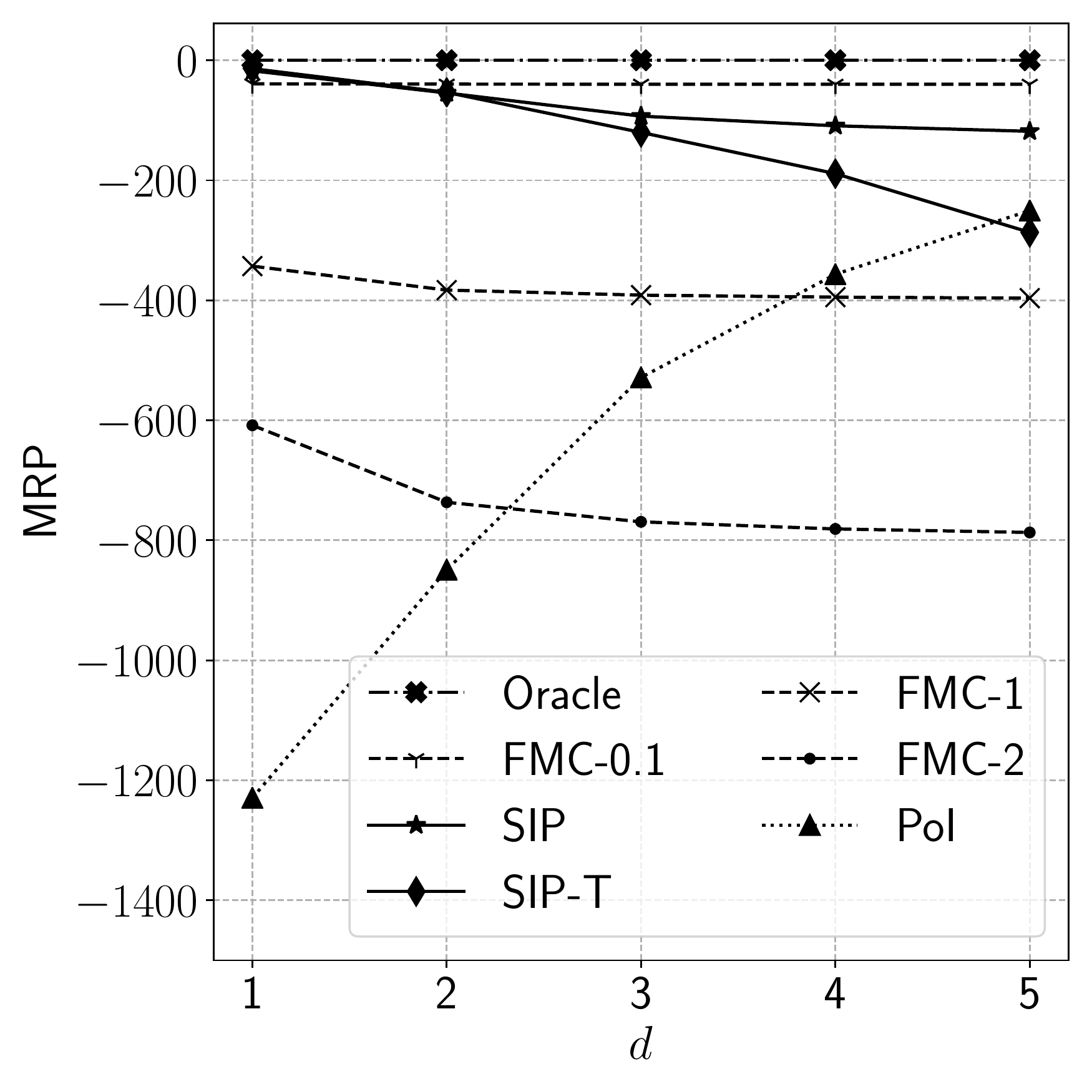}\label{fig:exp_vary_privacy_valuation_scale_linear_median_adjusted_payoff}}%
  
  \subfigure[Recall]{\scalebox{0.75}{
  \bgroup
        \def\arraystretch{1}%
        \addvbuffer[0pt 0pt]{\begin{tabular}[b]{l||r|r|r|r|r|}
 & \multicolumn{1}{c|}{\textbf{1}} & \multicolumn{1}{c|}{\textbf{2}} & \multicolumn{1}{c|}{\textbf{3}} & \multicolumn{1}{c|}{\textbf{4}} & \multicolumn{1}{c|}{\textbf{5}} \\ \hline \hline
\textbf{FMC-0.1} & 0.67 & 0.33 & 0 & 0 & 0 \\ \hline
\textbf{FMC-1} & 1 & 0.67 & 0.67 & 0.67 & 0.33 \\ \hline
\textbf{FMC-2} & 1 & 1 & 0.67 & 0.67 & 0.67 \\ \hline
\textbf{PoI} & 0.67 & 0.33 & 0 & 0 & 0 \\ \hline
\textbf{SIP} & 0.67 & 0.67 & 0.33 & 0.33 & 0.33 \\ \hline
\textbf{SIP-T} & 1 & 0.67 & 0.67 & 0.67 & 0.67 \\ \hline 
\end{tabular}\label{tbl:exp_vary_privacy_valuation_scale_recall}}
    \egroup}}
  \caption{The effect of the scale $\privscale$ of the privacy distributions}
  \label{fig:exp_vary_privacy_valuation_scale}
\end{figure}

Figure~\ref{fig:exp_vary_privacy_valuation_scale} shows the effect of the scale $\privscale$ of the users' privacy distributions. In all metrics, performance tends to decrease when the scale increases, because the FMC-based algorithms would obtain noisier data for the same price, and SIP and SIP-T would need to spend more to obtain the same level of accuracy of a data point. For PoI, a lower value of $\privscale$ make it more likely to buy a data point, since the expected profit gain would be higher, hence, it would spend more to buy data, resulting in a lower MRP; and vice versa.

\subsection{The effect of algorithmic parameters}
\subsubsection{The starting price $\price_0$}
\begin{figure}[htpb!]
  \centering
  \subfigure[ARP]{\includegraphics[scale=0.22]{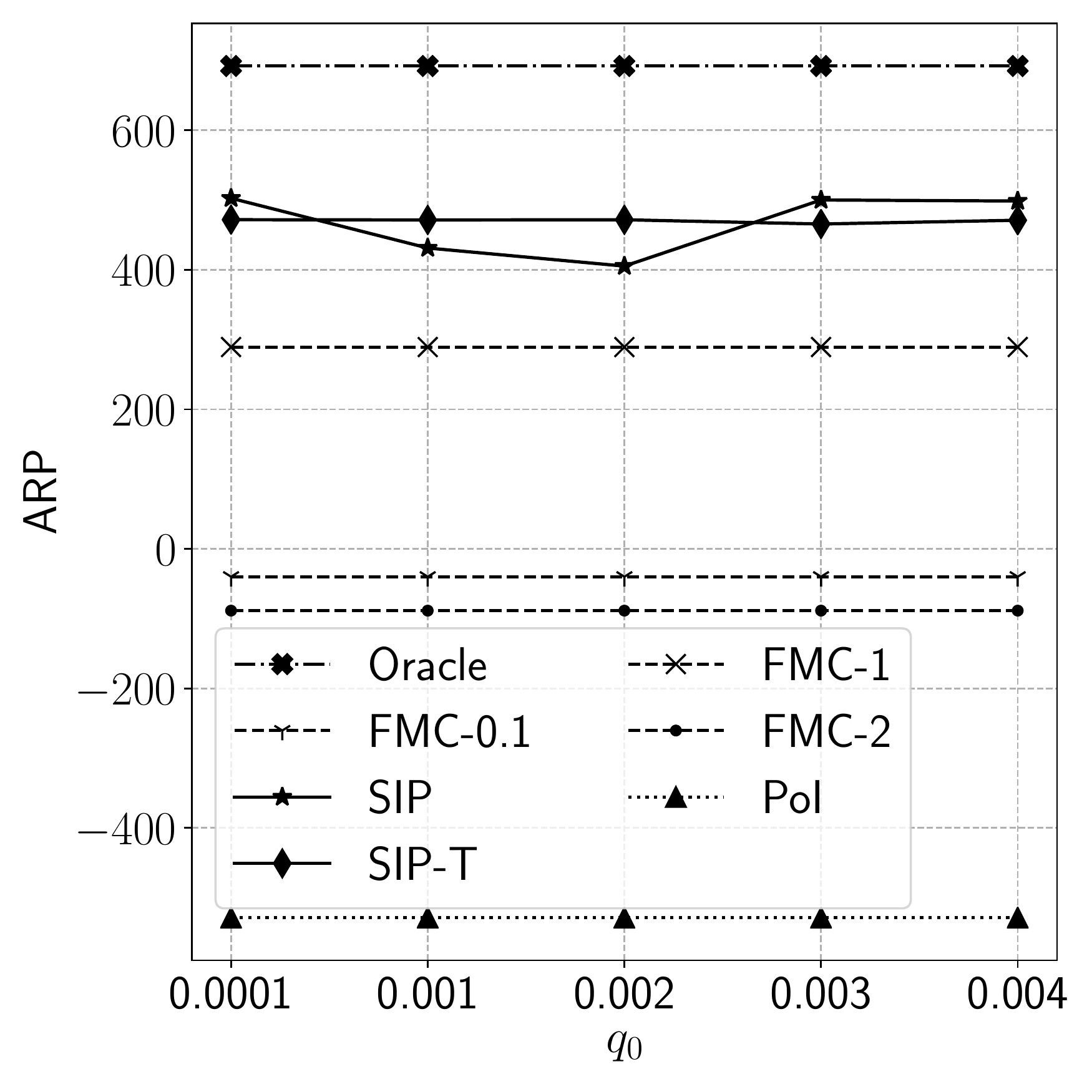}\label{fig:exp_vary_start_price_linear_avg_adjusted_payoff}}%
  \subfigure[MRP]{\includegraphics[scale=0.22]{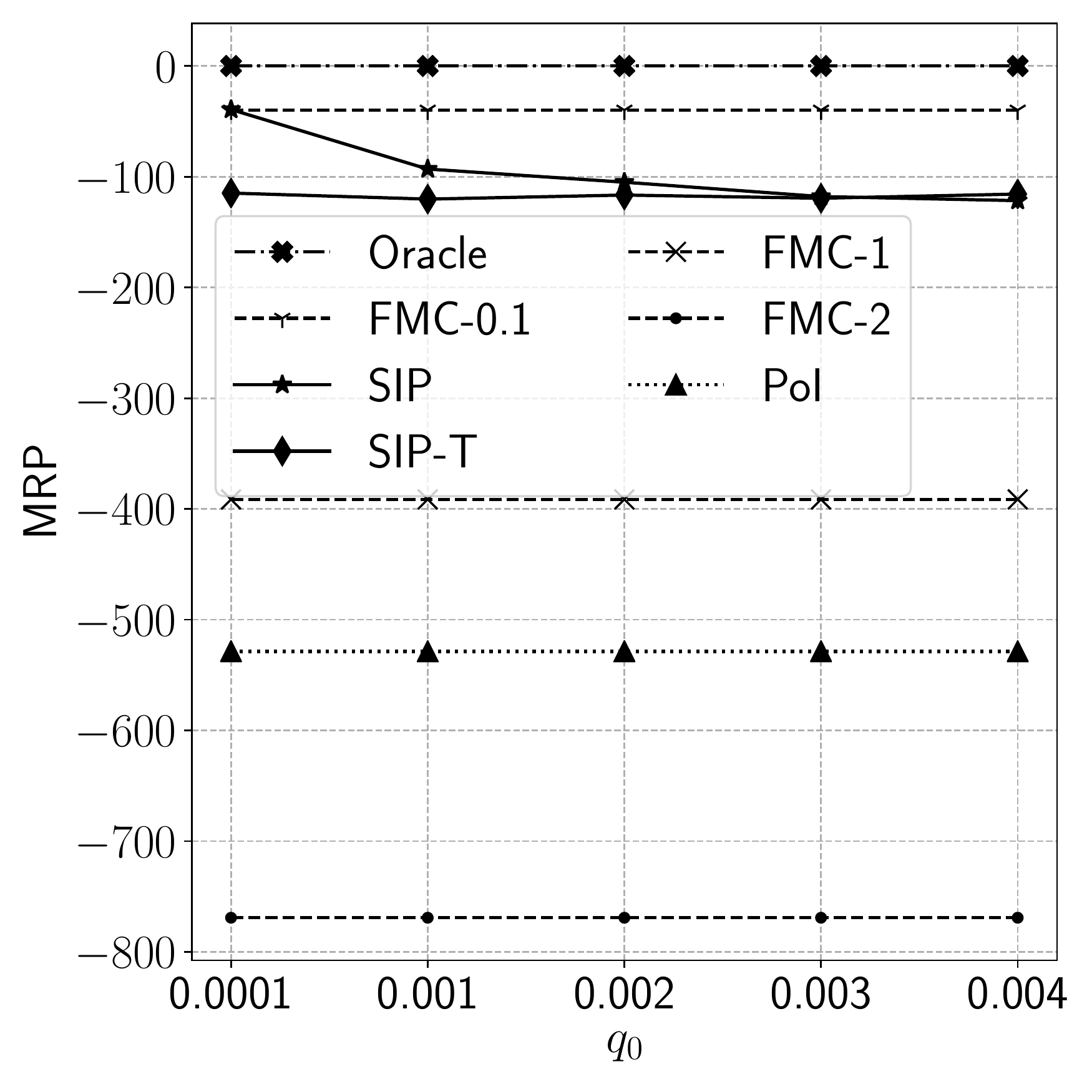}\label{fig:exp_vary_start_price_linear_median_adjusted_payoff}}%
  
  \subfigure[Recall]{\scalebox{0.75}{
  \bgroup
        \def\arraystretch{1}%
        \addvbuffer[0pt 0pt]{\begin{tabular}[b]{l||r|r|r|r|r|}
\textbf{FMC-0.1} & 0 & 0 & 0 & 0 & 0 \\ \hline
\textbf{FMC-1} & 0.67 & 0.67 & 0.67 & 0.67 & 0.67 \\ \hline
\textbf{FMC-2} & 0.67 & 0.67 & 0.67 & 0.67 & 0.67 \\ \hline
\textbf{PoI} & 0 & 0 & 0 & 0 & 0 \\ \hline
\textbf{SIP} & 0.33 & 0.33 & 0.33 & 0.67 & 0.67 \\ \hline
\textbf{SIP-T} & 0.67 & 0.67 & 0.67 & 0.67 & 0.67 \\ \hline 
\end{tabular}\label{tbl:exp_vary_start_price_recall}}
    \egroup}}
  \caption{The effect of the starting price $\price_0$}
  \label{fig:exp_vary_start_price}
\end{figure}

Figure~\ref{fig:exp_vary_start_price} shows the effect of the starting price $\price_0$. Because it is not preferable to spend a large amount during the pure exploration phase, the starting price is set to small values. 

While SIP-T uses the starting price, it eventually aims to distinguish whether a data point is inside or outside the target region. Therefore, as long as the starting price is relatively small, the change of the starting price does not have significant effect to the performance of SIP-T. However, for SIP, there is a small change in ARP since with a higher starting price, it can obtain more accurate data to make more accurate open/cancel decisions. However, it would decrease its MRP, which reflects the total cost of buying data. On the other hand, a very small value of $\price_0$ may result in too noisy data points at the beginning and can negatively effect the performance of SIP. 
Since SIP-T is more stable to the change of $\price_0$, it is more preferable if one is willing to spend more budget for buying data. 

\subsubsection{The price increment factor $\priceincreasefactor$}

Figure~\ref{fig:exp_vary_price_increment_factor} shows the effect of the price increment factor $\priceincreasefactor$. For both SIP and SIP-T, a higher value of $\priceincreasefactor$ means that they would skip calculating EIP of a potential price more often. This can decrease ARP but probably for different reasons: SIP may incorrectly stop buying a data point, resulting in a higher MRP because it might spend less; on the other hand, SIP-T may purchase data points at a too high price, resulting in a lower MRP because it might spend more.

With a smaller value of $\priceincreasefactor$, SIP may spend more on uninformative purchases, because the standard deviation of the next purchase may not be different enough from the current purchase, thus, increasing the total amount spent (i.e. a lower MRP). While both SIP and SIP-T spend more, SIP-T performs better than SIP with a small value of the factor $\priceincreasefactor$, because it can make more accurate open/cancel decisions. 

\begin{table}[htb]
    \centering
    
    \scalebox{0.8}{
  \bgroup
        \def\arraystretch{1}%
        \addvbuffer[0pt 0pt]{\begin{tabular}[b]{l||r|r|r|r|}
 & \multicolumn{1}{c|}{\textbf{1.5}} & \multicolumn{1}{c|}{\textbf{2}} & \multicolumn{1}{c|}{\textbf{3.5}} & \multicolumn{1}{c|}{\textbf{5}} \\ \hline \hline
\textbf{SIP} & 265 & 124 & 54 & 38 \\ \hline
\textbf{SIP-T} & 163 & 100 & 59 & 48 \\ \hline 
\end{tabular}}
    \egroup}
    \caption{The average execution time (in seconds) of SIP and SIP-T for different values of the price increment factor $\priceincreasefactor$}
    \label{tbl:exp_vary_price_increment_factor_exe_time}
\end{table}

A smaller value of $\priceincreasefactor$ would also result in a longer execution time of SIP and SIP-T since there would be more potential prices for which they need to calculate EIPs. Table~\ref{tbl:exp_vary_price_increment_factor_exe_time} shows the average execution time of SIP and SIP-T for different values of $\priceincreasefactor$.

\begin{figure}[htp!]
  \centering
  \subfigure[ARP]{\includegraphics[scale=0.22]{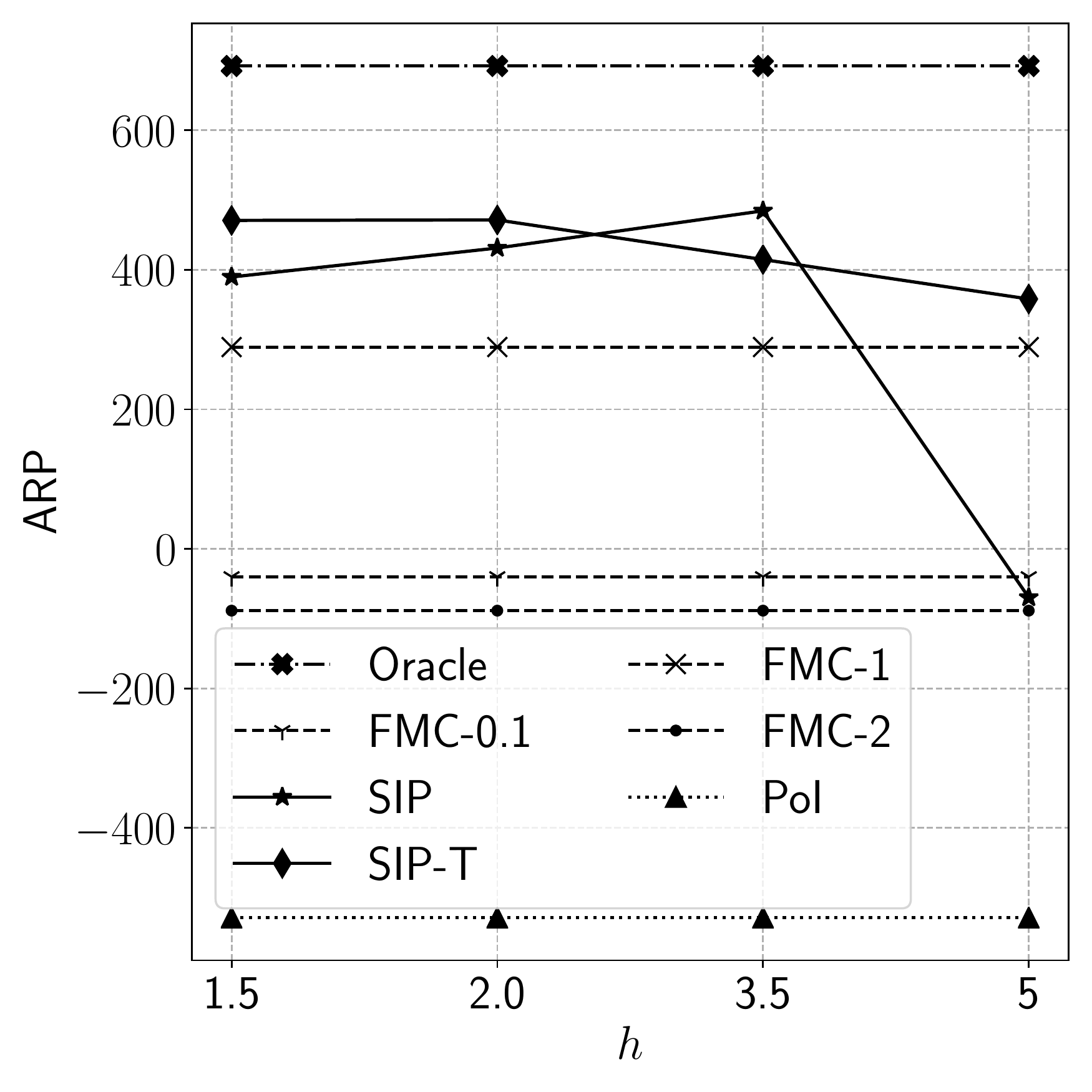}\label{fig:exp_vary_price_increment_factor_linear_avg_adjusted_payoff}}%
  \subfigure[MRP]{\includegraphics[scale=0.22]{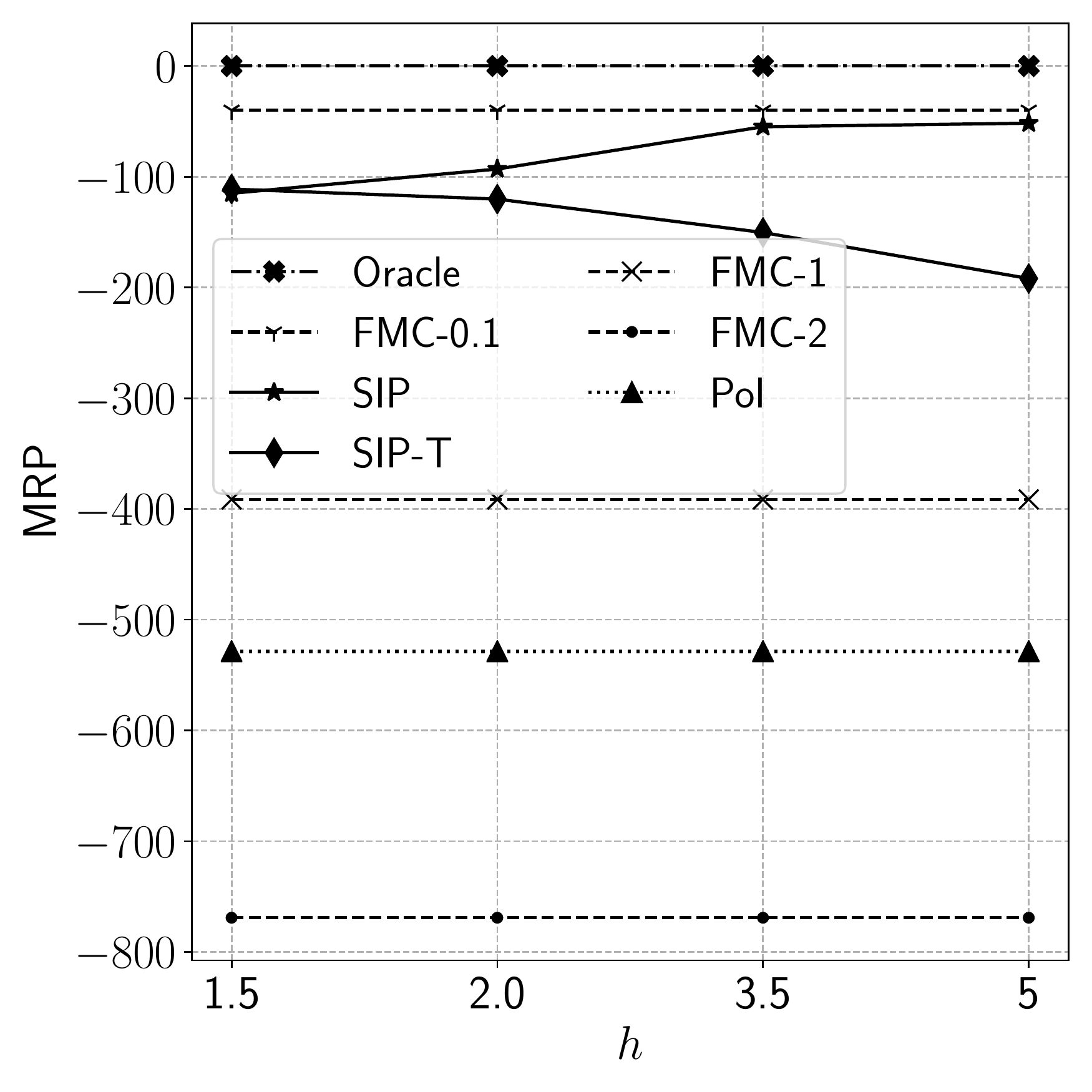}\label{fig:exp_vary_price_increment_factor_linear_median_adjusted_payoff}}%
  
  \subfigure[Recall]{\scalebox{0.75}{
  \bgroup
        \def\arraystretch{1}%
        \addvbuffer[0pt 0pt]{\begin{tabular}[b]{l||r|r|r|r|}
     & \multicolumn{1}{c|}{\textbf{1.5}} & \multicolumn{1}{c|}{\textbf{2}} & \multicolumn{1}{c|}{\textbf{3.5}} & \multicolumn{1}{c|}{\textbf{5}} \\ \hline \hline
    \textbf{FMC-0.1} & 0 & 0 & 0 & 0 \\ \hline
    \textbf{FMC-1} & 0.67 & 0.67 & 0.67 & 0.67 \\ \hline
    \textbf{FMC-2} & 0.67 & 0.67 & 0.67 & 0.67 \\ \hline
    \textbf{PoI} & 0 & 0 & 0 & 0 \\ \hline
    \textbf{SIP} & 0.33 & 0.33 & 0.33 & 0 \\ \hline
    \textbf{SIP-T} & 0.67 & 0.67 & 0.67 & 0.67 \\ \hline 
    \end{tabular}\label{tbl:exp_vary_price_increment_factor_recall}}
    \egroup}}
  \caption{The effect of the price increment factor $\priceincreasefactor$}
  \label{fig:exp_vary_price_increment_factor}
\end{figure}

\section{Discussions and Future Work}

To ease discussion, several simplifying assumptions were made for the buyer's profit maximization problem. For example, the privacy valuation was assumed to be known to the users or parameters of the profit model of the buyer were known; only a single snapshot of locations of users or only one buyer with one query was considered. Even with these simplifications, the buyer's profit maximization problem remains a challenging problem. For example, it can be seen as a particularly hard instance of POMDP. Future work should look at other scenarios that may illustrate new nuances to the central problem. In addition, while the proposed algorithms, SIP and SIP-T, consistently outperformed the baselines, other algorithms could also be explored to further optimize the buyer's decisions.

We also emphasize that the buyer's profit maximization problem is only one specific problem used to concretely demonstrate the broader problem: the spatial privacy pricing problem. For future work, we would explore other aspects of the spatial privacy pricing problem. For example, users' valuation of their data may change when they can observe selling and buying actions in a data marketplace, so we would explore the dynamic of users' valuation given the interactions in the marketplace. Other privacy preserving mechanism or other ways of ensuring privacy beyond adding noise (e.g. encryption) can also be employed. We would also study further the role of the marketplace or other types of queries of the buyers.

\section{Conclusions}
\label{sec:conclusions}
In a geo-marketplace, users can charge a price for their location data, and buyers can try to optimize their utility by making intelligent buying decisions. With \emph{spatial privacy pricing} we introduce the element of privacy, where a user can charge different prices depending on how much a particular data point would reveal. We illustrate this interplay between privacy, utility and price with an example scenario of a buyer who is considering opening a restaurant. The restaurant will only be profitable if there are enough people nearby. With this example, we formalize the privacy and pricing considerations of the seller as well as the buying and utility considerations of the buyer. 
The buyer's reasoning is captured as an incremental expectation maximization algorithm that accounts, in a principled way, for the points' uncertainty, prices and the anticipated profit.

Our formulation results in a new geospatial problem for optimizing a buyer's decision-making process. Using our formula for ``expected incremental profit'', we introduced two related algorithms for specifying which location points a buyer should buy at which prices. The algorithms look for the next best point to buy. Compared with five baseline algorithms, our SIP and SIP-T algorithms are able to better adapt to the locations and prices of user data.

To the best of our knowledge, this is the first research that considers the privacy, utility and price of location data in a unified framework. This is an important step for creating a real geo-marketplace.

\textbf{Acknowledgements}: This research has been funded in part by NSF grants IIS-1910950 and CNS-2027794, the USC Integrated Media Systems Center, and unrestricted cash gifts from Microsoft. Any opinions, findings, and conclusions or recommendations expressed in this material are those of the author(s) and do not necessarily reflect the views of any of the sponsors such as the NSF.

\bibliographystyle{ACM-Reference-Format}
\bibliography{bibliography}

\end{document}